\documentclass[prd,twocolumn,superscriptaddress,floatfix,amsmath,amssymb,amsfonts,nofootinbib,longbibliography]{revtex4-2}

\usepackage{float} 
\usepackage{scalerel}
\usepackage[normalem]{ulem}
\usepackage[english]{babel}
\usepackage{graphicx}
\usepackage{dcolumn}
\usepackage{bm}
\usepackage{blindtext}
\usepackage{verbatim}
\usepackage{relsize}
\usepackage{mathrsfs}
\usepackage{musicography}
\usepackage{amsmath}
\usepackage{blindtext}
\usepackage{cancel}
\usepackage{physics}
\usepackage{epstopdf}
\usepackage{mathtools}
\usepackage{blindtext}
\usepackage{tensor}
\usepackage{color}
\usepackage[usenames,dvipsnames]{pstricks}
\usepackage{epsfig}
\usepackage{pst-grad} 
\usepackage{pst-plot} 
\usepackage{hyperref}
\usepackage{verbatim}
\usepackage{slashed}
\usepackage{dsfont}
\usepackage[english]{babel}
\usepackage{amsmath,amssymb,amsthm}  

\newcommand{\mf}{\mathsf}

\newcommand{\ii}{\mathrm{i}}

\newcommand{\tc}[1]{\textsc{#1}}

\newcommand{\W}[2]{W_\textsc{#1}^{\scaleto{#2}{3.2pt}}}
\newcommand{\Wt}[1]{W_\tau^{\scaleto{#1}{3.2pt}}}
\newcommand{\Wmtc}[1]{\overline{W}^{\scaleto{#1}{3.2pt}}\!\!\!\!\!\!\!{}_{\scaleto{-}{3.2pt}\tau}}
\newcommand{\G}[2]{G_\textsc{#1}^{\scaleto{#2}{3.2pt}}}

\allowdisplaybreaks[1] 

\begin{document}

\title{Closed-form expressions for smeared bi-distributions of a massless scalar field: \\non-perturbative and asymptotic results in relativistic quantum information}

\author{T. Rick Perche}
\email{rick.perche@su.se}

\affiliation{Department of Applied Mathematics, University of Waterloo, Waterloo, Ontario, N2L 3G1, Canada}
\affiliation{Institute for Quantum Computing, University of Waterloo, Waterloo, Ontario, N2L 3G1, Canada}
\affiliation{Perimeter Institute for Theoretical Physics, Waterloo, Ontario, N2L 2Y5, Canada}

\begin{abstract}
   Using spacetime Gaussian test functions, we find closed-form expressions for the smeared Wightman function, Feynman propagator, retarded and advanced Green's functions, causal propagator, and symmetric propagator of a massless scalar field in the vacuum of Minkowski spacetime. We apply our results to localized quantum systems that interact with a quantum field in Gaussian spacetime regions and study different relativistic quantum information protocols. In the protocol of entanglement harvesting, we find a closed-form expression for the entanglement that can be acquired by probes that interact in Gaussian spacetime regions and obtain asymptotic results for the protocol. We also revisit the case of two gapless detectors and show that the detectors can become entangled if there is two-way signalling between their interaction regions, providing closed-form expressions for the detectors' final state.
\end{abstract}

\maketitle

\section{Introduction}

Interactions between quantum systems are fundamentally mediated by quantum fields. As the precision of our measurement devices and experimental techniques reach unprecedented levels, the effects of quantum fields in interactions between quantum systems become increasingly significant. In particular, there has been growing interest in understanding how the quantum features of mediating fields affect quantum information protocols~\cite{KojiCapacity,Simidzija_2020,Ahmadzadegan2021,quantClass,aspling2023high,alessio,LandulfoHappy,ourBMV}. These features can then be used to either enhance existing protocols~\cite{KojiCapacity,Ahmadzadegan2021}, or to allow entirely new processes to take place~\cite{Jonsson2,collectCalling,PRLHyugens2015,teleportation,Farming,Hotta2011,teleportation2014,nichoTeleport,Simidzija_2020,teleportExperiment}.

All features of quantum fields are encoded in the expected values of operators in quantum field theory (QFT). Moreover, for a specific class of states, namely, for quasifree states, all predictions of the theory are encoded in the field's two-point function, as well as other bi-distributions that can be built from it (such as the Feynman propagator). As such, if observers have access to the field in specific spacetime regions, it is enough to smear these bi-distributions in the corresponding regions to obtain the expected value of any field operator that the observers might have access to. One of the goals of this manuscript is to provide explicit closed-form expressions for the smeared field bi-distributions that observers have access to when they interact with the field in Gaussian regions of spacetime. 

Of special interest to us is the case where one has access to the field in localized regions of spacetime so that local operations can be applied to a quantum field. Local operations in QFT can be studied using probes that locally couple to the field. These probes can then be used to access the field's correlations between different spacetime regions~\cite{pipo,geometry}, and they can also be used to transmit classical and quantum information through the field~\cite{Jonsson2,collectCalling,Landulfo,ericksonCapacity,KojiCapacity,Ahmadzadegan2021}. Usually, the localized quantum systems used for these processes are modeled as particle detectors, introduced by Unruh~\cite{Unruh1976} in the context of black hole radiation, and later explored by DeWitt~\cite{DeWitt}. These models have since been referred to as Unruh-DeWitt (UDW) detectors, and have proven to be extremely valuable tools for studying properties of QFT~\cite{JormaHowThermal,waitUnruh,edunruh,JormaHawking,BenitoJormaUnruhState,bhDetectorsBTZ,jose,hectorjose}. Moreover, it has also been shown that results obtained from weakly interacting particle detectors directly translate to localized interactions entirely prescribed in terms of relativistic fields~\cite{QFTPD,FullyRelativisticEH}.

An example of a protocol in quantum field theory that can be implemented by UDW detectors is entanglement harvesting~\cite{Valentini1991,Reznik1,reznik2,Pozas-Kerstjens:2015}, which consists of using localized probes to extract pre-existing entanglement from a quantum field. Many studies of entanglement harvesting have been conducted considering different states of quantum fields in different spacetimes (see e.g.~\cite{Salton:2014jaa,Pozas2016,HarvestingSuperposed,Henderson2019,bandlimitedHarv2020,ampEntBH2020,ericksonNew,mutualInfoBH,SchwarzchildHarvestingWellDone,threeHarvesting2022,twist2022}, among others). Although entanglement harvesting has been studied in a plethora of scenarios, there are significant challenges in the explicit computations involved in the protocol. For instance, one must to consider weak interactions with the field in order to obtain explicit results using perturbation theory. Moreover, one has to resort to other approximations~\cite{Ng1} or numerical techniques to evaluate the field bi-distributions that yield the entanglement acquired by the probes. As a matter of fact, as of today, there are no closed-form expressions for the entanglement that can be harvested by finite-sized particle detectors in any spacetime. Our results for the field bi-distributions will then provide the first example of closed-form results in entanglement harvesting. As an application of these expressions, we will find asymptotic results for the protocol.

Another case of interest is when the particle detectors coupled to a quantum field have trivial internal dynamics (the case where the detectors do not have an internal energy gap). In this case, the interaction with the field can be computed non-perturbatively for arbitrary couplings in terms of the field's two-point functions. This has been explored in~\cite{Landulfo}, where the author considers one detector to interact first with the field, in order to transmit information to a second detector. Here we will generalize this result to arbitrary coupling regions and apply our closed-form expressions to study the conditions under which the detectors can become entangled with each other. We will show that when there is two-way causal contact between the probes, it is possible to get them entangled through the field, and we will explore explicit examples using our closed-form expressions. 

This manuscript is organized as follows. In Section~\ref{sec:bidist}, we review an algebraic formulation of quantum field theory, emphasizing the role of the relevant bi-distributions in the construction. In Section~\ref{sec:closed} we state some of the closed-form expressions found for these bi-distributions, leaving the more general results for Appendix~\ref{app}. In Section~\ref{sec:UDW} we briefly review the UDW model. In Section~\ref{sec:harvesting} we apply our closed-form expressions to the protocol of entanglement harvesting, and find asymptotic results for the entanglement that can be harvested by two spacelike separated particle detectors. In Section~\ref{sec:gapless} we study two gapless detectors with arbitrary interaction regions, extending the results of~\cite{Landulfo}. We also consider specific examples of gapless detectors interacting with a massless scalar field and show how to recover the previous results in the literature from our expressions. The conclusions of this work can be found in Section~\ref{sec:conclusions}.







\section{The bi-distributions that describe a quantum field theory}\label{sec:bidist}

There are numerous ways of formulating quantum field theory. For instance, one can formulate it in terms of canonical quantization, path integrals, or from an algebraic perspective~\cite{Wald2,fulling_1989,birrell_davies}. Even within the context of algebraic quantum field theory (AQFT), there are numerous different formulations that can be used, such as considering the Wightman axioms~\cite{Haag}, formulations in terms of *-algebras~\cite{Khavkine2015} or in terms of C*-algebras~\cite{kasiaFewsterIntro}. No matter which formulation is used, all of these have one thing in common: all expected values of the theory can be written in terms of n-point correlation functions of the field smeared in regions of spacetime. In particular, many situations of physical relevance, such as the case of quasifree states, allow one to compute all predictions of the theory in terms of two-point functions (or their time-ordered version). In this paper, we will be exclusively concerned with examples of quasifree states, where a characterization of the two-point functions is all that is required for a full description of the theory.

Mathematically, the two-point functions are bi-distributions in spacetime (which we denote here by $\mathcal{M}$). That is, they are functions of the form \mbox{$A: \mathcal{F}(\mathcal{M})\times \mathcal{F}(\mathcal{M}) \rightarrow \mathbb{C}$}, where $\mathcal{F}(\mathcal{M})$ is a suitable space of test functions where the distributions are defined. In the context of algebraic quantum field theory, it is usual to take $\mathcal{F}(\mathcal{M}) = C_0^\infty(\mathcal{M})$ as the set of compactly supported smooth functions in $\mathcal{M}$~\cite{kasiaFewsterIntro}, but $\mathcal{F}(\mathcal{M})$ can be often generalized (e.g. to Schartz space~\cite{Fredenhagen2015utr}). We write the action of a bi-distribution $A$ on test functions $f$ and $g$ as
\begin{equation}
    A(f,g) = \int \dd V \dd V' f(\mf x) g(\mf x') A(\mf x, \mf x'),
\end{equation}
where it is understood that whenever a distribution is evaluated at spacetime events $\mf x$ and $\mf x'$, $A(\mf x, \mf x')$ stands for the kernel of the distribution $A$. It is, however, not always possible to write a well-defined function $A(\mf x, \mf x')$ such that $A(f,g)$ takes the shape above, and in most cases $A(\mf x, \mf x')$ has a more general distributional interpretation as the kernel of $A(f,g)$
. Also notice that it is possible to apply a bi-distribution to only one of its arguments, resulting in a spacetime function. Indeed, if $A$ is a bi-distribution and $f\in \mathcal{F}(\mathcal{M})$, one can define the function
\begin{equation}
    Af(\mf x) = \int \dd V' A(\mf x, \mf x') f(\mf x')
\end{equation}
whenever the right-hand side of the equation above is well defined. Overall, if $A$ is a bi-distribution, we denote by $A(\mf x, \mf x')$ its integral Kernel, $A(f,g)$ its application to two test functions $f$ and $g$ and by $Af$ its application to a test function $f$, which results in a function in spacetime. This sets up the notation convention that will be used throughout the manuscript.

One example where bi-distributions are relevant is in the case of a real scalar Klein-Gordon field $\phi(\mf x)$ in a curved spacetime $\mathcal{M}$, which satisfies an equation of motion of the form
\begin{equation}\label{eq:eom}
    P \phi = (\nabla_\mu \nabla^\mu - V(\mf x)) \phi = 0,
\end{equation}
where $V(\mf x)$ is any smooth function in spacetime\footnote{for instance, a minimally coupled massive Klein-Gordon field corresponds to $V(\mf x) = m^2$}, and $P$ is defined as the differential operator that defines the homogeneous equations of motion $P\phi = 0$. The operator $P$ defines advanced and retarded Green's functions, $G_R$ and $G_A$, which can be used to find solutions to the non-homogeneous equation $P \phi = f$, where $f$ is a source function. The Green's functions satisfy the equations
\begin{align}\label{eq:PGd}
    P G_R f = f,\quad\quad
    P G_A f = f,
\end{align}
with the functions $G_Rf(\mf x)$ having support in the causal future of the support of $f$ and $G_Af(\mf x)$ having support in its causal past, corresponding to retarded and advanced propagation of a source $f$. An important property of the integral kernels of $G_R$ and $G_A$ is that
\begin{equation}
    G_R(\mf x, \mf x') = G_A(\mf x',\mf x),
\end{equation}
with the physical interpretation that propagation from $\mf x'$ to $\mf x$ in the forward time direction is equivalent to propagation from $\mf x'$ to $\mf x$ in the backwards time direction. The distributions $G_R$ and $G_A$ can also be used to build solutions to the homogeneous equation of motion through the causal propagator $E$, defined as
\begin{equation}\label{eq:E}
    E(\mf x,\mf x') = G_R(\mf x, \mf x') - G_A(\mf x, \mf x').
\end{equation}
Indeed, one can verify that $PEf = 0$ using \eqref{eq:PGd}. The causal propagator is also key in the definition of quantum field theories.

We will now briefly review the *-algebra formulation of a real scalar QFT, and highlight important properties of the bi-distributions that contain the information about the quantum field. In the context of this manuscript, a real scalar quantum field theory that satisfies the equation of motion~\eqref{eq:eom} will be defined by a $*$-algebra $\mathcal{A}$ which is generated by the identity $\openone$ and by products of operators of the form $\hat{\phi}(f)$, with $f\in C_0^\infty(\mathcal M)$, which satisfy the following properties
\begin{enumerate}
    \item $\hat{\phi}(\alpha f + g) = \alpha \hat{\phi}(f) + \hat{\phi}(g)$ for $\alpha\in \mathbb{C}$, $f,g\in C_0^\infty(\mathcal{M})$.
    \item $(\hat{\phi}(f))^\dagger = \hat{\phi}(f^*)$ for $f\in C_0^\infty(\mathcal{M})$, where $\dagger$ denotes the conjugation operation in the $*$-algebra and $*$ denotes complex conjugation.
    \item $\hat{\phi}(Pf) = 0$  for $f\in C_0^\infty(\mathcal{M})$.
    \item $[\hat{\phi}(f), \hat{\phi}(g)] = \ii E(f,g)$ for  $f,g\in C_0^\infty(\mathcal{M})$,
\end{enumerate}
where $E$ is the causal propagator, defined in~\eqref{eq:E}. In essence, property 1 ensures linearity of the association $f\mapsto \hat{\phi}(f)$, property 2 ensures that the field $\hat{\phi}$ is real, property 3 implements the equations of motion for the field $\hat{\phi}$, and property 4 defines the commutation structure of the algebra. The intuitive idea is that the algebra element $\hat{\phi}(f)$ can be seen as an operator-valued distribution in spacetime, which can formally be written as
\begin{equation}\label{eq:lie}
    \hat{\phi}(f) = \int \dd V f(\mf x) \hat{\phi}(\mf x),
\end{equation}
where $\hat{\phi}(\mf x)$ is in some way the `operator kernel' of this distribution. When we refer to `the quantum field' in this manuscript, we will usually mean the formal operator $\hat{\phi}(\mf x)$, with the understanding that well-defined algebra elements can be obtained by integrating it with respect to a suitable test function.

The formal expression of Eq.~\eqref{eq:lie} also allows one to define more general operators in the quantum field theory by considering distributional derivatives of the smeared field operators. For instance, the distribution $\nabla_\mu\hat{\phi}$, which acts in smooth compactly supported test vectors $j^\mu(\mf x)$, can be defined as
\begin{equation}
    \nabla \hat{\phi}(j) \coloneqq \hat{\phi}(-\nabla_\mu j^\mu) = \int \dd V (- \nabla_\mu j^\mu(\mf x))\hat{\phi}(\mf x),
\end{equation}
which allows one to formally recover an integrand of the form $j^\mu \nabla_\mu\hat{\phi}$ through integration by parts.

An important part of a quantum theory is its states. States in the context of AQFT are linear functionals \mbox{$\omega:\mathcal{A}\rightarrow \mathbb{C}$} which satisfy the conditions
\begin{enumerate}
    \item $\omega(\openone) = 1$.
    \item $\omega(\hat{A}^\dagger \hat{A}) \geq 0$ for all $\hat{A}\in \mathcal{A}$.
\end{enumerate}
The intuitive idea is that `states are mappings of observables into expected values', with the generalization to operators which are not necessarily self-adjoint. The analogy between a state $\omega$ in this context and a unit trace positive density operator $\hat{\rho}$ is $\omega(\hat{A}) = \text{tr}
 (\hat{\rho} \hat{A})$, whenever a representation $\hat{\rho}$ is well defined. In this sense, property 1 ensures normalization, and property 2 ensures positivity. From this perspective, a state $\omega$ is entirely defined by its value in arbitrary products of field operators. That is if one has access to $\omega(\hat{\phi}(f_1)...\hat{\phi}(f_n))$ for all possible $f_n$'s, then one has complete information about the state. This statement is equivalent to saying that a state is completely determined by its n-point functions, as one can always write
 \begin{equation}
     \omega(\hat{\phi}(f_1)...\hat{\phi}(f_n)) \!= \!\!\:\!\int \!\dd V_1 ... \dd V_n f_1(\mf x_1)...f_n(\mf x_n) W(\mf x_1,...,\mf x_n)
 \end{equation}
 for appropriate integral Kernels $W(\mf x_1,...,\mf x_2)$, which are precisely the n-point functions of the state. 

In this manuscript we will focus on the specific class of states that are quasifree, meaning that all their odd-point functions vanish, and all even-point functions can be computed from the two-point function $W(\mf x, \mf x')$ (or Wightman function) through Wick's theorem. This class of states is thus completely determined by their Wightman function, which showcases the importance of this specific bi-distribution in QFT. The (kernel of the) Wightman function can be decomposed into its real and imaginary parts using \mbox{$\hat{\phi}(f)\hat{\phi}(g) = \frac{1}{2}\{\hat{\phi}(f),\hat{\phi}(g)\}+\frac{1}{2}[\hat{\phi}(f),\hat{\phi}(g)]$}, so that 
\begin{align}
    W(f,g) &= \frac{1}{2}H(f,g) + \frac{\ii}{2}E(f,g),
\end{align}
where $E$ is the causal propagator and $H$ is the Hadamard function, $H(f,g) = \omega(\{\hat{\phi}(f),\hat{\phi}(g)\})$. Another important bi-distribution that can be obtained from the Wightman function is the Feynman propagator, whose kernel can be written as
\begin{equation}
    G_F(\mf x, \mf x') = \theta(t-t') W(\mf x, \mf x') + \theta(t'-t) W(\mf x', \mf x),
\end{equation}
where $\theta(u)$ denotes the Heaviside theta function and $t$ is any timelike coordinate. It can then be shown that
\begin{equation}\label{eq:GFHD}
    G_F(f,g)=  \frac{1}{2}H(f,g) + \frac{\ii}{2}\Delta(f,g),
\end{equation}
where $\Delta$ is the symmetric propagator:
\begin{equation}
    \Delta(f,g) = G_R(f,g) + G_A(f,g).
\end{equation}

The key points of this section are to set up the conventions that will be used throughout the manuscript and to convey the message that if one has knowledge of the distributions $W$ and $G_F$ for general functions $f$ and $g$, one can compute any expected values in a quasifree state. The shape of the functions $f$ and $g$ will then correspond to the regions where one has access to operators whose expected values one wishes to compute. One could then be led to believe that there is a rich literature with examples where the correlation functions of different states in different field theories are evaluated, and their properties are explored. Unfortunately, this is not the case. It is not common to find literature where the smeared two-point functions of quantum fields are evaluated in specific spacetime regions, and where these are used to infer properties of the field. The exceptions are studies of local operations in quantum field theory, usually utilizing quantum systems that couple to the field in localized spacetime regions. Even so, in most examples where this approach is considered, the evaluation of the relevant smeared two-point functions is mostly numerical. Overall, as far as the author is aware, no previous reference has closed-form expressions for all the smeared two-point functions described here smeared in any four dimensional spacetime regions.

\section{Closed-form expressions for Gaussian smeared bi-distributions in Minkowski spacetime}\label{sec:closed}

In this section, we will present closed form expressions for the bi-distributions discussed in Section~\ref{sec:bidist} evaluated at Gaussian test functions is spacetime. We will focus on the example of a real massless scalar field in Minkowski spacetime so that the differential operator $P$ in Eq. \eqref{eq:eom} reduces to $P = \nabla_\mu\nabla^\mu \equiv \Box$. Our goal will be to compute the distributions $W$ and $G_F$ evaluated at spacetime functions of the form
\begin{equation}\label{eq:f1f2}
    f(\mf x) = e^{- \frac{(t-t_0)^2}{2T^2}} e^{\ii \Omega t}\frac{e^{- \frac{|\bm x-\bm L|^2}{2\sigma^2}}}{(2\pi \sigma^2)^\frac{3}{2}} \equiv \chi(t) F(\bm x),
\end{equation}
which depend on the free parameters $(T,t_0,\Omega, \sigma, \bm L)$. The parameters $T$ and $\sigma$ correspond to the standard deviations in time and in space of $f(\mf x)$, and define the region where the function is effectively non-zero\footnote{At this stage it must be clear that functions of the form of Eq.~\eqref{eq:f1f2} do not belong to the space $C_0^\infty(\mathcal{M})$. Nevertheless, these functions are localized, in the sense that for a fixed small $\epsilon>0$, the region $|f(\mf x)|>\epsilon$ is finite. The idea is that the value $|f(\mf x)|$ defines ``how much one has access to the quantum field'' at the event $\mf x$.}. The parameters $t_0$ and $\bm L$ define the center of the effective support of $f(\mf x)$, and $\Omega$ can be seen as a shift in frequency through the Fourier transform.

In particular, in this section we will compute $H(f_1,f_2)$, $E(f_1,f_2)$, $G_R(f_1,f_2)$, $G_A(f_1,f_2)$, $\Delta(f_1,f_2)$ in the Minkowski vacuum, with the parameter choices $(T,0,\Omega, \sigma, \bm 0)$ for $f_1$ and $(T,t_0,\Omega, \sigma, \bm L)$ for $f_2$. Expressions with more general parameters can be found in Appendix~\ref{app}. Notice that the bi-distributions $H(f_1,f_2)$, $E(f_1,f_2)$ and $G_A(f_1,f_2)$ are enough to evaluate $W(f_1,f_2)$, $G_R(f_1,f_2)$, $G_F(f_1,f_2)$, as well as $\Delta(f_1,f_2)$. As far as the author is aware, the closed-form analytical expressions\footnote{Notice that the term ``closed-form analytical expression'' is ambiguous, as any integral expression can be considered to be closed-form. In this manuscript, we will use ``closed-form analytical expression'' to mean an expression in terms of special functions (as listed, for instance, in~\cite{gradshteyn}), with no need for any external dummy index or integration variable. This excludes any integrals or infinite series of functions.}
for all these bi-distributions in the Minkowski vacuum have not yet been found in the literature.

In order to perform the computations, we start by noting that the Wightman function in the Minkowski vacuum can be represented as
\begin{equation}
    W(\mf x, \mf x') = \frac{1}{(2\pi)^3}\int \frac{\dd^3 \bm k}{2|\bm k|} e^{- \ii |\bm k| (t-t')+\ii \bm k\cdot (\bm x - \bm x')},
\end{equation}
which allows one to recast the smeared Wightman function in terms of Fourier transforms, as has been done many times in the literature (see e.g.~\cite{Pozas-Kerstjens:2015,Pozas2016,ericksonNew}),
\begin{equation}\label{eq:WABk}
    W(f_1,f_2) = \frac{1}{(2\pi)^3}\int \frac{\dd^3\bm k}{2 |\bm k|} \tilde{F}_1(\bm k)\tilde{F}_2(-\bm k)\tilde{\chi}_2(|\bm k|)\tilde{\chi}_1(-|\bm k|),
\end{equation}
where $\tilde{F}(\bm k)$ and $\tilde{\chi}(\omega)$ denote the Fourier transforms
\begin{align}
    \tilde{F}(\bm k) = \int \dd^3 \bm x f(\bm x) e^{\ii \bm k \cdot \bm x},\quad
    \tilde{\chi}(\omega) = \int \dd t\chi(t) e^{\ii \omega t}.
\end{align}
We can then compute the integral of Eq.~\eqref{eq:WABk}. From its real part, we find
\begin{align}
    H(f_1,f_2) = \frac{T^2e^{-\Omega^2T^2}e^{\ii \Omega t_0}}{4 \sqrt{\pi} |\bm L| \sqrt{T^2+\sigma^2}}&\Big(e^{- \frac{(|\bm L| + t_0)^2}{4(T^2 + \sigma^2)}}\text{erfi}\left(\tfrac{|\bm L| +t_0}{2\sqrt{T^2 + \sigma^2}}\right)\nonumber\\
    &\!\!\!\!\!\!\!\!\!\!\!+e^{- \frac{(|\bm L| - t_0)^2}{4(T^2 + \sigma^2)}}\text{erfi}\left(\tfrac{|\bm L| -t_0}{2\sqrt{T^2 + \sigma^2}}\right)\Big),\label{eq:H}
\end{align}
and from the imaginary part of $W(f_1,f_2)$, we obtain
\begin{equation}\label{eq:EAB}
    E(f_1,f_2) = \frac{T^2e^{-\Omega^2T^2}e^{\ii \Omega t_0}}{4 \sqrt{\pi} |\bm L| \sqrt{T^2+\sigma^2}}\left(e^{-\frac{(|\bm L| + t_0)^2}{4(T^2 + \sigma^2)}}\!-\!e^{-\frac{(|\bm L| - t_0)^2}{4(T^2 + \sigma^2)}}\right)\!.
\end{equation}
These results have been found in~\cite{Pozas-Kerstjens:2015} with different conventions, but they are not sufficient to compute $G_A(f_1,f_2)$, $G_R(f_1,f_2)$, $\Delta(f_1,f_2)$, and, most importantly, $G_F(f_1,f_2)$. It is however possible to obtain all of these from $G_A(f_1,f_2)$, together with $H(f_1,f_2)$ and $E(f_1,f_2)$. While the integration in terms of Fourier transforms is non-trivial for the retarded and advanced Green's functions, we can switch strategies, and work in spacetime in order to solve these integrals. In inertial coordinates in Minkowski spacetime the retarded and advanced Green's functions of the operator $\Box$ can be written as
\begin{equation}
    G_R(\mf x, \mf x') = -\frac{1}{4\pi|\bm x - \bm x'|}\delta(t'-t+|\bm x - \bm x'|),\label{eq:GR}
\end{equation}
\begin{equation}
    G_A(\mf x, \mf x') = -\frac{1}{4\pi|\bm x - \bm x'|}\delta(t'-t-|\bm x - \bm x'|).\label{eq:GA}
\end{equation}
We can then perform the integration in spacetime (see Appendix~\ref{app}) to find
\begin{widetext}
\begin{align}
    G_R(f_1,f_2) &= -\frac{T^2e^{-\Omega^2T^2}e^{\ii \Omega t_0}}{8 \sqrt{\pi} |\bm L| \sqrt{T^2+\sigma^2}}\left(e^{-\frac{(|\bm L| + t_0)^2}{4 (T^2 + \sigma^2)}}\left(-1 + \text{erf}\left(\tfrac{|\bm L| T^2 - t_0 \sigma^2}{2 T \sigma \sqrt{T^2 + \sigma^2}}\right)\right)+e^{-\frac{(|\bm L| - t_0)^2}{4 (T^2 + \sigma^2)}}\left(1 + \text{erf}\left(\tfrac{|\bm L| T^2 + t_0 \sigma^2}{2 T \sigma \sqrt{T^2 + \sigma^2}}\right)\right)\right),\nonumber\\
    G_A(f_1,f_2) &= -\frac{T^2e^{-\Omega^2T^2}e^{\ii \Omega t_0}}{8 \sqrt{\pi} |\bm L| \sqrt{T^2+\sigma^2}}\left(e^{-\frac{(|\bm L| + t_0)^2}{4 (T^2 + \sigma^2)}}\left(1 + \text{erf}\left(\tfrac{|\bm L| T^2 - t_0 \sigma^2}{2 T \sigma \sqrt{T^2 + \sigma^2}}\right)\right)+e^{-\frac{(|\bm L| - t_0)^2}{4 (T^2 + \sigma^2)}}\left(-1 + \text{erf}\left(\tfrac{|\bm L| T^2 + t_0 \sigma^2}{2 T \sigma \sqrt{T^2 + \sigma^2}}\right)\right)\right),
\end{align}
which is compatible with Eq.~\eqref{eq:EAB} from $E = G_R-G_A$. We also find
\begin{equation}
    \Delta(f_1,f_2) = -\frac{T^2e^{-\Omega^2T^2}e^{\ii \Omega t_0}}{4 \sqrt{\pi} |\bm L| \sqrt{T^2+\sigma^2}}\left(e^{-\frac{(|\bm L| + t_0)^2}{4 (T^2 + \sigma^2)}}\text{erf}\left(\frac{|\bm L| T^2 - t_0 \sigma^2}{2 T \sigma \sqrt{T^2 + \sigma^2}}\right)+e^{-\frac{(|\bm L| - t_0)^2}{4 (T^2 + \sigma^2)}} \text{erf}\left(\frac{|\bm L| T^2 + t_0 \sigma^2}{2 T \sigma \sqrt{T^2 + \sigma^2}}\right)\right),
\end{equation}
\end{widetext}
from which the Feynman propagator $G_F(f_1,f_2)$ can be computed using Eq.~\eqref{eq:GFHD}. These expressions are particularly important in the context of entanglement harvesting, as we discuss in more detail in Section~\ref{sec:harvesting}, but have not been found in closed-form in previous works, which have instead used numerical integration to evaluate the Feynman propagator for different parameters (see e.g.~\cite{Salton:2014jaa,Pozas2016,HarvestingSuperposed,Henderson2019,bandlimitedHarv2020,ampEntBH2020,ericksonNew,mutualInfoBH,SchwarzchildHarvestingWellDone,threeHarvesting2022,twist2022,cisco2023harvesting}). It is also worth mentioning that in~\cite{Ng1} an expression was found for the limit $\sigma\to 0$ or our expressions. For generalizations of these expressions for different parameters $(T,t_0,\Omega, \sigma, \bm L)$ and for more general field operators coupled in these regions, see Appendix~\ref{app}.

As we will see in the next section, the results above are enough to compute all predictions that observers that couple to the field amplitude in regions defined by functions of the form $|f(\mf x)|$ have access to.

\section{Locally Probing a Quantum Field: Unruh-DeWitt Detectors}\label{sec:UDW}

In this section we will review the two-level UDW particle detector~\cite{Unruh1976,DeWitt} coupled to a scalar field. This is probably the simplest model of particle detector, and consists of a two-level system linearly coupled to a real scalar field. For all purposes, one can think of a UDW detector as a qubit that undergoes a given timelike trajectory $\mf z(\tau)$ in spacetime, and that interacts with a quantum field locally around its trajectory. Given how simple this model is, it can be perhaps surprising that it has been show to capture fundamental features of the interactions of any localized quantum system with a quantum field~\cite{generalPD}, such as the interactions of atoms and electromagnetism~\cite{Pozas2016,richard}, the interactions of nucleons with neutrinos~\cite{neutrinos,antiparticles,carol}, as well as the interactions of quantum systems with linearized quantum gravity~\cite{pitelli,boris}. Overall, UDW detectors capture the essential way in which localized quantum systems can be used to access information from quantum fields and implement local operations in QFT.

In order to define the two-level UDW detector, we assume that the qubit has a free Hamiltonian with energy gap $\Omega\geq0$, which promotes time evolution with respect to the proper time of the trajectory $\mf z(\tau)$ where the qubit is defined. The free Hamiltonian of the qubit can be written as
\begin{equation}
    \hat{H}_\tc{d} = \Omega \hat{\sigma}^+ \hat{\sigma}^-,
\end{equation}
where $\hat{\sigma}^\pm$ are the standard $SU(2)$ ladder operators. The Hamiltonian above then defines the ground and excited states $\{\ket{g},\ket{e}\}$ for the two-level system, so that $\hat{\sigma}^+ \ket{g} = \ket{e}$ and $\hat{\sigma}^-\ket{g} = 0$. 

The next step is to prescribe the interaction of the detector with a scalar quantum field $\hat{\phi}(\mf x)$. We assume that the interaction happens in a region associated with the support of a spacetime smearing function $\Lambda(\mf x)$, defined locally around the trajectory $\mf z(\tau)$. The value of $\Lambda(\mf x)$ at each $\mf x$ defines how much access the qubit has to the quantum field at each event, and here we will assume that the function $\Lambda(\mf x)$ is localized in spacetime, allowing the detector to access the field only in a localized region. The detector couples to the field through its monopole moment, which we prescribe as $\hat{\mu}(\tau) = e^{\ii \Omega \tau}\hat{\sigma}^+ + e^{- \ii \Omega \tau}\hat{\sigma}^-$, where free evolution with respect to $\hat{H}_\tc{d}$ has already been incorporated. The interaction Hamiltonian can then be written at the level of a scalar spacetime density $\hat{h}_I(\mf x)$~\cite{us}, given by
\begin{align}
    \hat{h}_I(\mf x) &= \lambda  \Lambda(\mf x)\hat{\mu}(\tau) \hat{\phi}(\mf x)\\
    &= \lambda \left(\Lambda^+(\mf x)\hat{\sigma}^+ + \Lambda^-(\mf x)\hat{\sigma}^-\right) \hat{\phi}(\mf x),
\end{align}
where $\lambda$ is a coupling constant and we defined
\begin{equation}
    \Lambda^\pm(\mf x) = \Lambda(\mf x) e^{\pm \ii \Omega \tau}.
\end{equation}
In the equation above, the time parameter $\tau$ stands for the Fermi normal coordinate time around the trajectory $\mf z(\tau)$ (see~\cite{us,us2,generalPD} and references therein for details). 

In order to see exactly how the detector's final state depends on the field, we proceed to compute the qubit's final state after the interaction. The time evolution operator can be computed from the interaction Hamiltonian density $\hat{h}_I(\mf x)$ as
\begin{equation}\label{eq:UI1d}
    \hat{U}_I = \mathcal{T}_\tau\exp\left(-\ii \int \dd V \hat{h}_I(\mf x)\right),
\end{equation}
where $\mathcal{T}_\tau$ denotes time ordering with respect to the time parameter $\tau$. It is important to remark that the time evolution operator of Eq.~\eqref{eq:UI1d} explicitly depends on the time parameter $\tau$ that is used to prescribe the qubit's internal degree of freedom. This has been discussed in detail in~\cite{us2}, and the reason for this breakdown of covariance is that the interaction Hamiltonian density fails to satisfy the microcausality condition. That is, $\hat{h}_I(\mf x)$ does not commute with itself at spacelike separated points, due to the fact that the monopole $\hat{\mu}(\tau)$ couples to $\hat{\phi}(\mf x)$ at each surface $\tau = \text{const.}$ at multiple spacelike separated points simultaneously. Although this could bring a fair concern to whether the UDW model should still be used in relativistic scenarios, the model's regimes of validity are well understood~\cite{eduardoOld,us,us2,PipoFTL,mariaPipoNew}, which allows one to use particle detectors to study relativistic quantum field theories in a plethora of scenarios.

In order to compute the time evolution operator explicitly, we use the Dyson series as a power expansion in the coupling constant, so that $\hat{U}_I$ can be written, to second order in $\lambda$, as
\begin{equation}
    \hat{U}_I = \openone + \hat{U}_I^{(1)} + \hat{U}_I^{(2)} + \mathcal{O}(\lambda^3),
\end{equation}
where
\begin{align}
    \hat{U}_I^{(1)} &= -\ii \int \dd V \hat{h}_I(\mf x),\\
    \hat{U}_I^{(2)} &= - \int \dd V \dd V' \theta(\tau - \tau')\hat{h}_I(\mf x)\hat{h}_I(\mf x').
\end{align}
We assume that the initial state of the detector is uncorrelated with the field, so that the initial state of the detector-field system can be written as $\hat{\rho}_0 = \hat{\rho}_{\tc{d},0}\otimes \hat{\rho}_\omega$, where $\hat{\rho}_{\tc{d},0}$ is an arbitrary qubit state, and $\hat{\rho}_\omega$ is the density operator in a given Hilbert space representation of a quasifree state of the field, $\omega$. As such, the state $\omega$ is entirely determined by the value of the field's Wightman function, $W(\mf x, \mf x') = \text{tr}(\hat{\rho}_\omega \hat{\phi}(\mf x)\hat{\phi}(\mf x')) = \omega(\hat{\phi}(\mf x) \hat{\phi}(\mf x'))$, and we assume from this point on that all field bi-distributions are evaluated at the state $\omega$. 

We are interested in the final state of the detector, only. That is, we work under the assumption that one only has access to the detector's degrees of freedom. Operationally, this is implemented by tracing over the quantum field, so that the detector state after the interaction is given by
\begin{align}\label{eq:Dyson1d}
    &\hat{\rho}_\tc{d} = \tr_\phi(\hat{U}_I \hat{\rho}_0\hat{U}_I^\dagger)\\
    & = \hat{\rho}_{\tc{d},0} + \tr_\phi(\hat{U}_I^{(1)} \hat{\rho}_0\hat{U}_I^{(1)\dagger})+ \tr_\phi(\hat{U}_I^{(2)} \hat{\rho}_0)+ \tr_\phi(\hat{\rho}_0\hat{U}_I^{(2)\dagger})\nonumber\\
    &\:\:\:\:\:\:\:\:\:\:\:\:\:\:\:\:\:\:\:\:\:\:\:\:\:\:\:\:\:\:\:\:\:\:\:\:\:\:\:\:\:\:\:\:\:\:\:\:\:\:\:\:\:\:\:\:\:\:\:\:\:\:\:\:\:\:\:\:\:\:\:\:\:\:\:\:\:\:\:\:\:\:\:\:\:\:\:\:\:\:+\mathcal{O}(\lambda^4).\nonumber
\end{align}
In order to compute the terms above, we first define the following bi-distributions, which are explicitly dependent on the time parameter $\tau$:
\begin{align}
    W_\tau(\mf x, \mf x') &= \theta(\tau - \tau')W(\mf x,\mf x'),\\
    W\!\!{}_{\scaleto{-}{3.2pt}\tau}(\mf x, \mf x') &= \theta(\tau' - \tau)W(\mf x',\mf x).
\end{align}
The two-point functions $W_\tau$ and $ W\!\!{}_{\scaleto{-}{3.2pt}\tau}$ can then be used to generate all the bi-distributions discussed in Section~\ref{sec:bidist}. For instance, one can express the Wightman function and the Feynman propagator as
\begin{align}\label{eq:propWt}
    W = W_\tau + W^*\!\!\!\!\!{}_{\scaleto{-}{3.2pt}\tau},\quad G_F = W_\tau + W\!\!{}_{\scaleto{-}{3.2pt}\tau},
\end{align}
where we remark that $(W_\tau(f,g))^* = W\!\!{}_{\scaleto{-}{3.2pt}\tau}(g^*,f^*)$. The advantage of using $W_\tau$ and $ W\!\!{}_{\scaleto{-}{3.2pt}\tau}$ in the context of particle detectors is that these depend explicitly on the time parameter used to define the model itself, and will naturally show up in the Dyson series for the detector's final state. Indeed, we can write
\begin{align}
    &\tr_\phi(\hat{U}_I^{(2)} \hat{\rho}_0) =\!- \!\!\int \!\dd V \dd V' \theta(\tau - \tau') \tr_\phi(\hat{h}_I(\mf x)\hat{h}_I(\mf x')\hat{\rho}_\omega)\hat{\rho}_{\tc{d},0}\nonumber\\
    &= - \lambda^2 \left(W_\tau(\Lambda^+,\Lambda^-)\hat{\sigma}^+\hat{\sigma}^- + W_\tau(\Lambda^-,\Lambda^+)\hat{\sigma}^- \hat{\sigma}^+\right)\hat{\rho}_{\tc{d},0},
\end{align}
and, analogously,
\begin{align}
     &\tr_\phi(\hat{\rho}_0\hat{U}_I^{(2)\dagger} ) = \tr_\phi(\hat{U}_I^{(2)} \hat{\rho}_0)^\dagger \\
    &= - \lambda^2\hat{\rho}_{\tc{d},0}\left(W^*\!\!\!\!\!{}_{\scaleto{-}{3.2pt}\tau}(\Lambda^+,\Lambda^-)\hat{\sigma}^+\hat{\sigma}^- + W^*\!\!\!\!\!{}_{\scaleto{-}{3.2pt}\tau}(\Lambda^-,\Lambda^+)\hat{\sigma}^- \hat{\sigma}^+\right).\nonumber
\end{align}
The term $\tr_\phi(\hat{U}_I^{(1)} \hat{\rho}_0\hat{U}_I^{(1)\dagger})$ in Eq.~\eqref{eq:Dyson1d} does not depend explicitly on the time ordering, so that it is independent of the parameter $\tau$,
\begin{align}
    \tr_\phi(&\hat{U}_I^{(1)} \hat{\rho}_0\hat{U}_I^{(1)\dagger}) = \int \dd V \dd V' \tr_\phi(\hat{h}_I(\mf x)\hat{\rho}_{0}\hat{h}_I(\mf x'))\\
    &= \int \dd V \dd V' \Lambda(\mf x) \Lambda(\mf x')W(\mf x',\mf x)\hat{\mu}(\tau)\hat{\rho}_{\tc{d},0}\hat{\mu}(\tau')\nonumber\\
    &= \lambda^2\Big(W(\Lambda^+,\Lambda^-)\hat{\sigma}^+\hat{\rho}_{\tc{d},0}\hat{\sigma}^+ + W(\Lambda^-,\Lambda^+)\hat{\sigma}^+\hat{\rho}_{\tc{d},0}\hat{\sigma}^-\nonumber\\
    &\:\:\:\:\:+ W(\Lambda^+,\Lambda^-)\hat{\sigma}^-\hat{\rho}_{\tc{d},0} \hat{\sigma}^+ + W(\Lambda^-,\Lambda^-)\hat{\sigma}^- \hat{\rho}_{\tc{d},0} \hat{\sigma}^- \Big)\nonumber
\end{align}
Combining the results above, we then obtain the final state of the detector, to leading order:
\begin{align}\label{eq:rhod1d}
    \hat{\rho}_\tc{d} = \hat{\rho}_{\tc{d},0} + &(\W{}{++}\hat{\sigma}^+\hat{\rho}_{\tc{d},0}\hat{\sigma}^+ + \W{}{-+}\hat{\sigma}^+\hat{\rho}_{\tc{d},0}\hat{\sigma}^- \\
    &+ \W{}{+-}\hat{\sigma}^-\hat{\rho}_{\tc{d},0} \hat{\sigma}^+ + \W{}{--} \hat{\sigma}^- \hat{\rho}_{\tc{d},0} \hat{\sigma}^-\nonumber\\
    & -\Wt{+-}\hat{\sigma}^+ \hat{\sigma}^- \hat{\rho}_{\tc{d},0} - \Wt{-+}\hat{\sigma}^- \hat{\sigma}^+ \hat{\rho}_{\tc{d},0}\nonumber\\
    &- \Wmtc{+-}\hat{\rho}_{\tc{d},0}\hat{\sigma}^+\hat{\sigma}^- - \Wmtc{-+}\hat{\rho}_{\tc{d},0}\hat{\sigma}^-\hat{\sigma}^+) \nonumber\\
    &\:\:\:\:\:\:\:\:\:\:\:\:\:\:\:\:\:\:\:\:\:\:\:\:\:\:\:\:\:\:\:\:\:\:\:\:\:\:\:\:\:\:\:\:\:\:\:\:\:\:\:\:+\mathcal{O}(\lambda^4),\nonumber
\end{align}
where we denote
\begin{align}
    \W{}{\pm\mp} &= \lambda^2W(\Lambda^\pm,\Lambda^\mp),\\ 
    \Wt{\pm\mp} &= \lambda^2W_\tau(\Lambda^\pm,\Lambda^\mp),\\ 
    \Wmtc{\pm\mp} &= \lambda^2 W^*_{\scaleto{-}{3.2pt}\tau}(\Lambda^\pm,\Lambda^\mp).
\end{align}
From Eq.~\eqref{eq:rhod1d} one can see that the leading order result for $\hat{\rho}_\tc{d}$ will in general depend on the time parameter $\tau$ due to the dependence on the bi-distributions $W_\tau$ and $W_{\!\scaleto{-}{3.2pt}\tau}$. Indeed, it is only when $[\hat{\rho}_{\tc{d},0},\hat{\sigma}^+\hat{\sigma}^-] = [\hat{\rho}_{\tc{d},0},\hat{\sigma}^-\hat{\sigma}^+] = 0$ that one has a result independent of the time parameter $\tau$, which is one of the key results of~\cite{us2}. This can also be seen from Eq. \eqref{eq:rhod1d} by noting that $\W{}{-+} = \Wt{-+}+\Wmtc{-+}$ (see Eq. \eqref{eq:propWt}).

For concreteness, we specialize the result to the case where the detector starts in the ground state, with \mbox{$\hat{\rho}_{\tc{d},0} = \ket{g}\!\!\bra{g} = \hat{\sigma}^-\hat{\sigma}^+$}. In this case we obtain
\begin{equation}\label{eq:detGstate}
    \hat{\rho}_{\tc{d}} = (1 - \W{}{-+})\hat{\sigma}^-\hat{\sigma}^+ + \W{}{-+}\hat{\sigma}^+\hat{\sigma}^- + \mathcal{O}(\lambda^4),
\end{equation}
where $\W{}{-+}$ is the leading order excitation probability of the state, and we see that it is determined by the field's Wightman function evaluated at the spacetime smearing function $\Lambda(\mf x)$ with an added complex phase due to the detector's free time evolution. In the case where the detector interacts with the Minkowski vacuum in a Gaussian spacetime region, one can find a closed-form expression for $\W{}{-+}$ (as has been done in~\cite{Pozas-Kerstjens:2015,ericksonNew}, among others). For completeness, we state the result for $\W{}{-+}$ in the case where
\begin{equation}
    \Lambda(\mf x) = e^{- \frac{t^2}{2T^2}}\frac{e^{- \frac{|\bm x|^2}{2\sigma^2}}}{(2\pi \sigma^2)^\frac{3}{2}}.
\end{equation}
We obtain
\begin{equation}\label{eq:Wpm}
    \W{}{-+} = \frac{\lambda^2}{4\pi} \frac{e^{- \Omega^2 T^2}}{\alpha^2}\left(1 - \frac{\sqrt{\pi}\Omega T}{\alpha}e^{\frac{\Omega^2T^2}{\alpha^2}}\text{erfc}\left(\frac{\Omega T}{\alpha}\right)\right),
\end{equation}
where we defined $\alpha = \sqrt{1 + \sigma^2/T^2}$. From Eq. \eqref{eq:Wpm} one can explore properties of the interaction of the detector with the field. For instance, one can see that the vacuum excitation probability vanishes in the limit of large interaction times. This is often interpreted as the fact that it is only possible to see ``virtual field excitations'' for finite interaction times. 

\section{Entanglement harvesting and asymptotic results}\label{sec:harvesting}

In this section we will consider the protocol of entanglement harvesting, where two UDW detectors couple to a quantum field in an attempt to extract entanglement previously existent in the field. At this stage, there is a vast literature in entanglement harvesting, with examples in both flat~\cite{Salton:2014jaa,Pozas2016,HarvestingSuperposed,Henderson2019,bandlimitedHarv2020} and curved spacetimes~\cite{ampEntBH2020,ericksonNew,mutualInfoBH,SchwarzchildHarvestingWellDone}. Nevertheless, the lack of closed-form expressions for field bi-distributions makes it so that one usually has to employ numerical methods to evaluate the integrals that allow one to compute the final state of the detectors involved in the protocol. The only exception was in~\cite{Ng1}, where closed-form expressions were found in the case of two point-like inertial UDW detectors in Minkowski spacetime interacting with a massless scalar field. In this section we will use the results of Section~\ref{sec:closed} to generalize the expressions of~\cite{Ng1} to smeared detectors. We will also use our closed-form expressions to find asymptotic results for the entanglement that can be extracted from a real massless quantum field by a pair of detectors.

In order to implement the protocol of entanglement harvesting, we consider two UDW detectors defined along trajectories $\mf z_\tc{a}(\tau_\tc{a})$ and $\mf z_\tc{b}(\tau_\tc{b})$ with energy gaps $\Omega_\tc{a}$ and $\Omega_\tc{b}$ coupled to a massless scalar field $\hat{\phi}(\mf x)$ according to the interaction Hamiltonian density
\begin{equation}
    \hat{h}_I(\mf x) = \lambda \left(\hat{\mu}_\tc{a}(\tau_\tc{a})\Lambda_\tc{a}(\mf x) + \hat{\mu}_\tc{b}(\tau_\tc{b})\Lambda_\tc{b}(\mf x) \right)\hat{\phi}(\mf x).
\end{equation}
Here $\tau_\tc{a}$ and $\tau_\tc{b}$ represent the Fermi normal coordinate times associated with each trajectory. We also assume that the supports of $\Lambda_\tc{a}(\mf x)$ and $\Lambda_\tc{b}(\mf x)$ do not extend past the region where the Fermi normal coordinates are defined.  

When considering two detectors, it is again possible to use the Dyson series to express the time evolution operator as a power series in the coupling constant $\lambda$. The computations are similar to those of Section~\ref{sec:UDW} for an initial state for the detectors-field system of the form $\hat{\rho}_0 = \hat{\rho}_{\tc{ab},0}\otimes \hat{\rho}_\omega$, with $\hat{\rho}_\omega$ being the representation of a quasifree state $\omega$. One finds that the leading order final state of the two detectors after tracing the quantum field can be written as
\begin{align}
    \hat{\rho}_\tc{ab} = \hat{\rho}_{\tc{ab},0} + \lambda^2 \!\!\!\sum_{{\substack{{}_{s_1,s_2 = \pm}\\\tc{i},\tc{j} = \tc{a},\tc{b}}}}\!\! &W(\Lambda^{s_1}_\tc{i},\Lambda^{s_2}_\tc{j})\hat{\sigma}^{s_2}_\tc{j}\hat{\rho}_{\tc{ab},0}\hat{\sigma}_\tc{i}^{s_1} \\[-12pt]
    &\:\:\!-\! \left(W_\tau(\Lambda^{s_1}_\tc{i},\Lambda^{s_2}_\tc{j})\hat{\sigma}^{s_1}_\tc{i}\hat{\sigma}_\tc{j}^{s_2}\hat{\rho}_{\tc{ab},0} + \text{H.c.}\right) \nonumber\\[2pt]
    &\:\:\:\:\:\:\:\:\:\:\:\:\:\:\:\:\:\:\:\:\:\:\:\:\:\:\:\:\:\:\:\:\:\:\:\:\:\:\:\:\:\:\:\:\:\:\:\:\:\:+\mathcal{O}(\lambda^4),\nonumber
\end{align}
where here $\tau$ is the time parameter used to prescribe the time evolution in the model. Unfortunately, the equation above is not particularly insightful. In order to analyze the final state of the detectors, let us consider a specific choice of initial state for $\hat{\rho}_{\tc{ab},0}$. As it is usually considered in protocols of entanglement harvesting~\cite{Pozas-Kerstjens:2015,Pozas2016,HarvestingSuperposed,ericksonNew,hectorMass}, we will initialize both detectors in their ground state, considering the uncorrelated state $\hat{\rho}_{\tc{ab},0} = \ket{g_\tc{a}}\!\!\bra{g_\tc{a}}\otimes \ket{g_{\tc{b}}}\!\!\bra{g_\tc{b}} = \hat{\sigma}_\tc{a}^- \hat{\sigma}_\tc{a}^+ \hat{\sigma}_\tc{b}^- \hat{\sigma}_\tc{b}^+$. In this setup it is simpler to write the final density operator in matrix form in the basis $\{\ket{g_\tc{a}g_\tc{b}},\ket{e_\tc{a}g_\tc{b}},\ket{g_\tc{a}e_\tc{b}},\ket{e_\tc{a}e_\tc{b}}\}$. We obtain
\begin{align}\label{eq:rhoABharvest}
    \hat{\rho}_\tc{ab} = &\begin{pmatrix}
                            1 - (\W{aa}{-+}+\W{bb}{-+}) & 0 & 0 & -  (\G{ab}{++})^*\\
                            0 & \W{bb}{-+} &  (\W{ab}{-+})^* & 0 \\
                            0 &  \W{ab}{-+} &  \W{aa}{-+} & 0\\
                            -  \G{ab}{++} & 0 & 0 & 0
                            \end{pmatrix}\nonumber\\
                        & \:\:\:\:\:\:\:\:\:\:\:\:\:\:\:\:\:\:\:\:\:\:\:\:\:\:\:\:\:\:\:\:\:\:\:\:\:\:\:\:\:\:\:\:\:\:\:\:\:\:\:\:\:\:\:\:\:\:\:\:\:\:\:\:\:\:\:\:\:\:+ \mathcal{O}(\lambda^4),
\end{align}
and we denote
\begin{align}
    \W{ij}{\pm \mp} &= \lambda^2 W(\Lambda_\tc{i}^\pm,\Lambda_\tc{j}^\mp),&&& \G{ij}{\pm \mp} &= \lambda^2 G_F(\Lambda_\tc{i}^\pm,\Lambda_\tc{j}^\mp).
\end{align}
Importantly, as far as the author is aware, a closed-form expression for the function $G_F(\Lambda_\tc{i},\Lambda_\tc{j})$ smeared in four-dimensional spacetime has not been obtained in any previous work. This has important consequences to one's ability to estimate the optimal parameters that allow the detectors to harvest entanglement from the vacuum of a quantum field.

In order to check the entanglement present in the detectors' final state, we will use the negativity~\cite{WootersAlone,VidalNegativity,horodeckiReview} as an entanglement quantifier. The negativity is a faithful entanglement quantifier for bipartite systems of qubits, and is defined as the sum of the absolute value of the negative eigenvalues of the partial transpose of a density operator. For the specific density operator of Eq. \eqref{eq:rhoABharvest}, the leading order negativity can be written as
\begin{align}
    \mathcal{N}(\hat{\rho}_\tc{ab}) = &\text{max}\left(0,\sqrt{|\G{ab}{++}|^2 + \left(\tfrac{\W{aa}{-+} - \W{bb}{-+}}{2}\right)^2} - \tfrac{\W{aa}{-+}+\W{bb}{-+}}{2}\right)\nonumber\\
     &\:\:\:\:\:\:\:\:\:\:\:\:\:\:\:\:\:\:\:\:\:\:\:\:\:\:\:\:\:\:\:\:\:\:\:\:\:\:\:\:\:\:\:\:\:\:\:\:\:\:\:\:\:\:\:\:\:\:\:+ \mathcal{O}(\lambda^4).
\end{align}
Overall, the negativity is a competition between the non-local term $\G{ab}{++}$ and the local vacuum noise terms $\W{aa}{-+}$ and $\W{bb}{-+}$. This competition becomes even more explicit in the case of where the excitation probabilities are the same, $\W{aa}{-+} = \W{bb}{-+} = \mathcal{L}$, when the negativity reduces to
\begin{equation}
    \mathcal{N}(\hat{\rho}_{\tc{ab}}) = \max(0,|\G{ab}{++}|-  \mathcal{L}).
\end{equation}

At this stage it is important to mention that there are two physically distinguished processes that can allow the detectors to become entangled. One possibility is that the detectors become entangled after exchanging information through the field, in which case the field merely propagates quantum information from one detector to another. The other case is when the detectors extract previously existing entanglement from the field's state itself. The differences between these have been explicitly studied in~\cite{ericksonNew} and the role of the field's quantum degrees of freedom in the protocol has been discussed in~\cite{quantClass}. 

It is possible to quantify how much of the entanglement acquired by the detectors is due to communication and how much of it is actual entanglement harvested from the field by looking at the real and imaginary parts of the Feynman propagator $G_F = \frac{1}{2}H + \frac{\ii}{2}\Delta$ (Eq.~\eqref{eq:GFHD}). The symmetric propagator $\Delta(\mf x, \mf x')$ represents the causal interaction between the detectors and is responsible for exchanging information between them. An evidence of the fact that $\Delta(\mf x, \mf x')$ does not contain any information about the pre-existing correlations of the field is that this propagator is independent of the field's state. On the other hand, the Hadamard function $H(\mf x, \mf x')$ contains all the state dependence of the propagator $G_F(\mf x, \mf x')$, and as such, it contains the information about the non-local correlations in the quantum field. In order to successfully harvest entanglement from a QFT, one looks for situations where the contribution of the symmetric propagator $\Delta(\mf x, \mf x')$ is negligible, and where the contribution due to the Hadamard term $H(\mf x, \mf x')$ is dominating. One way of ensuring this is by considering spacelike separated interaction regions, where $\Delta(\Lambda^+_\tc{a},\Lambda^+_\tc{b}) = 0$. However, even when the interactions are not completely causally separated, one can achieve entanglement harvesting if $\mathcal{N}(\hat{\rho}_\tc{d}) >0$ and $ \frac{1}{2}|\Delta(\Lambda^+_\tc{a},\Lambda^+_\tc{b})|\ll \mathcal{N}(\hat{\rho}_\tc{d})$.

In order to consider an explicit example of entanglement harvesting, we will consider the case where the detectors interact with a massless scalar field in Minkowski spacetime, with the following spacetime smearing functions
\begin{align}
    \Lambda_\tc{a}(\mf x) &= e^{- \frac{t^2}{2 T^2}} \frac{e^{-\frac{|\bm x|^2}{2 \sigma^2}}}{(2\pi \sigma^2)^\frac{3}{2}},\label{eq:LA}\\
    \Lambda_\tc{b}(\mf x) &= e^{- \frac{(t-t_0)^2}{2 T^2}} \frac{e^{-\frac{|\bm x-\bm L|^2}{2 \sigma^2}}}{(2\pi \sigma^2)^\frac{3}{2}}.\label{eq:LB}
\end{align}
This choice defines the interaction regions of the detectors to be spacetime Gaussians of spatial width $\sigma$ and effective time duration controlled by the parameter $T$. The interaction regions are shifted in space by $\bm L$ and in time by $t_0$ with respect to the inertial frame $(t,\bm x)$. Due to the fact that the spacetime smearing functions of Eqs.~\eqref{eq:LA} and~\eqref{eq:LB} differ only by spacetime translations, we find that $\W{aa}{-+} = \W{bb}{-+} = \mathcal{L}$. This scenario has been studied multiple times in the literature (see e.g.~\cite{Pozas-Kerstjens:2015,Pozas2016,ericksonNew,hectorMass,Ng1}), however, it was only in the $\sigma \to 0$ limit that analytical results were found for the relevant smeared bi-distributions necessary to compute the negativity. While the $\mathcal{L}$ term can be evaluated analytically and is given by Eq. \eqref{eq:Wpm}, in order to numerically evaluate $\G{ab}{++}$ effectively, one usually writes, in momentum space~\cite{Pozas-Kerstjens:2015},
\begin{align}\label{eq:GABk}
    \G{ab}{++} = \frac{\lambda^2 T^2e^{\ii \Omega t_0}}{4 \pi }\!\int &\dd |\bm k|\,|\bm k| e^{- |\bm k|^2 \sigma^2} e^{-(\Omega^2 + |\bm k|^2)T^2} \text{sinc}(|\bm k||\bm L|)\nonumber\\
    &\times \Bigg(e^{-\ii |\bm k| t_0}\text{erfc}\left(\ii |\bm k|T -\frac{t_0}{2T}\right)\nonumber\\&\:\:\:\:\:\:\:\:\:\:\:\:\:\:\:+e^{\ii |\bm k| t_0}\text{erfc}\left(\ii |\bm k|T +\frac{t_0}{2T}\right)\Bigg),
\end{align}
which up to this point could not be solved in terms of elementary functions.

On the other hand, the results that we have from Section~\ref{sec:closed} give the exact value of the relevant propagators in this case. We find
\begin{widetext}
\begin{align}
    \G{ab}{++} =
    \frac{\lambda^2 T^2e^{-\Omega^2T^2}e^{\ii \Omega t_0}}{8 \sqrt{\pi} |\bm L| \sqrt{T^2+\sigma^2}}\Bigg(&e^{- \frac{(|\bm L| + t_0)^2}{4(T^2 + \sigma^2)}}\text{erfi}\left(\frac{|\bm L| +t_0}{2\sqrt{T^2 + \sigma^2}}\right)+e^{- \frac{(|\bm L| - t_0)^2}{4(T^2 + \sigma^2)}}\text{erfi}\left(\frac{|\bm L| -t_0}{2\sqrt{T^2 + \sigma^2}}\right)\nonumber\\
    &-\ii\left(e^{-\frac{(|\bm L| + t_0)^2}{4 (T^2 + \sigma^2)}}\text{erf}\left(\frac{|\bm L|T^2 - t_0\sigma^2 }{2 T \sigma\sqrt{T^{2} + \sigma^{2}}}\right)+e^{-\frac{(|\bm L| - t_0)^2}{4 (T^2 + \sigma^2)}} \text{erf}\left(\frac{|\bm L|T^2 + t_0\sigma^2 }{2 T \sigma\sqrt{T^{2} + \sigma^{2}}}\right)\right)\Bigg).\label{eq:GAB}
\end{align}
We have also checked that the numerical integration of Eq.~\eqref{eq:GABk} matches the results of Eq.~\eqref{eq:GAB} for numerous parameters $|\bm L|$, $t_0$, $T$, $\sigma$ and $\Omega$. Combining the equation above with Eq.~\eqref{eq:Wpm}, one then finds a closed-form analytical expression for the negativity. In particular, for the case where there is no time separation between the interactions, $t_0 = 0$, we obtain, with $\alpha = \sqrt{1 + \sigma^2/T^2}$,
\begin{equation}
    \mathcal{N}(\hat{\rho}_\tc{ab}) = \max\left(0,\frac{\lambda^2 e^{-\Omega^2 T^2}}{4\pi \alpha^2 } \left(\sqrt{\pi}\alpha e^{- \frac{|\bm L|^2}{4\alpha^2T^2}}\frac{T}{|\bm L| }\sqrt{\text{erf}\left(\frac{|\bm L|}{2 \alpha \sigma}\right)^2 + \text{erfi}\left(\frac{|\bm L|}{2\alpha T}\right)^2}  + \frac{\sqrt{\pi}\Omega T}{\alpha}e^{\frac{\Omega^2T^2}{\alpha^2}}\text{erfc}\left(\frac{\Omega T}{\alpha}\right)- 1 \right)\right),\label{eq:neg}
\end{equation}
\end{widetext}
while the imaginary part of the Feynman propagator at $t_0 = 0$ reads
\begin{equation}\label{eq:DAB}
    \tfrac{1}{2}|\Delta(\Lambda_\tc{a}^+,\Lambda_\tc{b}^+)| = \frac{e^{-\Omega^2 T^2}}{4\alpha \sqrt{\pi}} \frac{T}{|\bm L|}e^{- \frac{|\bm L|^2}{4\alpha^2T^2}}\text{erf}\left(\frac{|\bm L|}{2 \alpha \sigma}\right).
\end{equation}
Eq.~\eqref{eq:DAB} can then be used to estimate the signalling between the detectors, so that we are looking for situations in which $\mathcal{N}(\hat{\rho}_\tc{ab})>0$ and $\frac{1}{2}|\Delta_{\tc{ab}}^{\scaleto{++}{3.2pt}}|\ll\mathcal{N}(\hat{\rho}_\tc{ab})$. From the expression above, one also confirms the exponential decay of the signalling between two Gaussian detectors which was seen in~\cite{mariaPipoNew}. We also see that the $\text{erf}$ term in Eq. \eqref{eq:neg} comes exclusively from the signalling between the detectors.

\begin{figure}[h!]
    \centering
    \includegraphics[width=8.6cm]{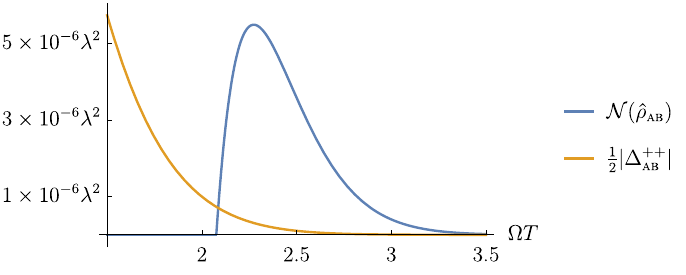}
    \caption{The negativity and signalling contribution for two detectors interacting with a massless scalar field in Gaussian spacetime regions separated by $|\bm L| = 5 T$, with detector sizes $\sigma = 0.01T$.}
    \label{fig:n}
\end{figure}

In Fig.~\ref{fig:n}, we plot the negativity and the signalling contribution as a function of the detectors' energy gap for $\sigma = 0.01 T$ when the detectors are separated by a distance of $|\bm L| = 5 T$. We find that the negativity becomes an order of magnitude larger than the signalling from the moment at which it peaks, and this ratio continues to increase with $\Omega T$.

Notice that the results of Eq.~\eqref{eq:GAB} can also be used to determine the previously unsolved integral of Eq.~\eqref{eq:GABk} in closed-form. For instance, comparing Eqs.~\eqref{eq:GABk} and~\eqref{eq:GAB} at $t_0 = 0$, we find
\begin{align}\label{eq:newInt}
    \int_0^\infty \!\!\dd r\, e^{-(\gamma^2 + \sigma^2)r^2} &\text{sin}(r\ell) \,\text{erfi}\left(\gamma r\right)\\
    &= \frac{e^{- \frac{\ell^2}{4(\gamma^2 + \sigma^2)}}}{ \sqrt{\gamma^2+\sigma^2}}\frac{\sqrt{\pi}}{2} \,\text{erf}\left(\tfrac{\ell \gamma}{2 |\sigma| \sqrt{\gamma^{2} + \sigma^{2}}}\right),\nonumber
\end{align}
which, as far as the author is aware, has not been solved in closed-form in previous literature. The result above is valid for $\ell,\gamma,\sigma \in \mathbb{R}$ with $|\gamma|<|\sigma|$, but it also seems to work for complex $\gamma$ and $\ell$, provided that $|\gamma|<|\sigma|$, so that the integral on the left-hand side converges.

We now turn to the study of asymptotic results in entanglement harvesting in the weak coupling regime. Having the analytical expression for the negativity allows one to estimate the behaviour of the entanglement that can be harvested by detectors that couple to a massless field in the limit of large $|\bm L|$. This is done by considering the asymptotic expansion of Eq.~\eqref{eq:GAB}. One finds
\begin{equation}
    |\G{ab}{++}| = \frac{\lambda^2e^{- \Omega^2 T^2}}{2\pi} \left(\frac{T^2}{|\bm L|^2} + \frac{2T^2(T^2+\sigma^2)}{|\bm L|^4}+ \mathcal{O}\left(|\bm L|^{-6}\right)\right)\!.
\end{equation}
Notice, however, that the term above must be larger than the local vacuum excitations of the detectors in order to ensure that any entanglement can be harvested at all. The vacuum noise is independent of $|\bm L|$, but, as has been shown in~\cite{hectorMass}, there is always a large enough value of the energy gap $\Omega$ which allows the Gaussian detectors to harvest entanglement. In order to see which value of $\Omega$ maximizes the negativity, we differentiate Eq.~\eqref{eq:neg} assuming $\sigma\ll T$, and set the result to zero, which yields the same asymptotic result for $\Omega$ found in~\cite{hectorMass}: $\Omega T \sim | \bm L|/(2T)$. This allows us to evaluate the negativity at said energy gap, and to consider the asymptotic limit of the resulting expression. We find that for $\Omega T \sim | \bm L|/(2T)$,
\begin{equation}
    \mathcal{N}(\hat{\rho}_{\tc{ab}}) = \frac{4\lambda^2 e^{-\frac{|\bm L|^2}{4 T^2}}}{\pi }\frac{T^4}{|\bm L|^4} + \mathcal{O}\left(\frac{T^6}{|\bm L|^6}e^{-\frac{|\bm L|^2}{4 T^2}}\right).
\end{equation}
We can then conclude the first asymptotic result regarding entanglement harvesting: the maximum entanglement that can be harvested by a pair of inertial detectors that couple to a massless field in Gaussian spacetime regions decays as $\ell^{-4}e^{-\ell^2/4}$ in the limit of large separations, where $\ell =  |\bm L|/T$. In this limit, of course, the communication between the detectors is negligible, as they are too far away to signal to each other. Indeed, the signalling term, in this case, decays as $e^{- \ell^2/2}$.

Another interesting asymptotic result regarding the entanglement acquired by the detectors can be obtained in the case of very large interaction times, $T\to \infty$. However, in this case, the main contribution for the entanglement acquired by the detectors is given by the signalling term, so that the setup does not properly configure entanglement harvesting. In any case, one obtains
\begin{equation}
    \mathcal{N}(\hat{\rho}_\tc{ab}) = \frac{\lambda^2e^{-\Omega^2 T^2}e^{- \frac{|\bm L|^2}{4T^2}}}{4\sqrt{\pi}} \frac{T} {|\bm L|}\text{erf}\left(\tfrac{|\bm L|}{2 \sigma}\right) + \mathcal{O}\left(\frac{e^{-\frac{|\bm L|^2}{4T^2}}}{T}\right),
\end{equation}
and the vacuum excitation probabilities are also negligible in this limit, as they behave as $1/(\Omega T)^2$ in the limit of large $T$ (for $\Omega>0$). Although the asymptotic limit above cannot be used for estimating entanglement harvesting, it is useful as an estimate of how entangled two systems that interact with a massless quantum field could become in the limit of large interaction times. Notice that larger values of the energy gap will result in less entangled systems. The physical interpretation for this fact is that the larger $\Omega$ is, the more energy it costs to change the state of either qubit from the ground state.

\section{Non-perturbative results for gapless detectors}\label{sec:gapless}

In this section we will show non-perturbative results for gapless detectors ($\Omega = 0$). This case has been studied in the past, for instance in~\cite{Landulfo}, however, the results of~\cite{Landulfo} are only valid when the interaction regions of two detectors are separated by a Cauchy surface. Our goal in this section is to generalize these results to arbitrary interaction regions for the detectors and to employ the results of Section~\ref{sec:closed} to analyze specific examples. We will focus on the case of a single gapless detector in Subsection~\ref{sub:1gapless}, and on the case of two gapless detectors in Subsection~\ref{sub:2gapless}. We compare our results with previous literature in Subsection~\ref{sub:comparison}.

Before going through the computations with gapless detectors, let us discuss the physical interpretation of this limit of the UDW model. One of the applications of particle detectors is to provide a definition of the concept of particles measured by an observer. This is done by considering a detector in a given state of motion, and by associating excitations of the detector with the detection of field quanta. However, it is necessary to have an energy gap in order to claim that energy from the field was absorbed, and thus, to make statements about ``detected particles''. Gapless detectors do not share this property, and therefore cannot be used to discuss particle absorption/emission. Nevertheless, gapless detectors can be used to extract field correlations more precisely than gapped detectors can.

Intuitively, one can think of a UDW detector as a spin-1/2 system with an energy gap $\Omega$ which is put to interact with a scalar field. The energy gap $\Omega$ in the case of a spin system is often the result of the application of an external (classical) magnetic field. In terms of the Bloch sphere, the effect of the energy gap is to add a constant rotation around the axes of the magnetic field. While this constant rotation is in place, the qubit is then put to interact with a quantum field, and the field fluctuations effectively generate another axis of rotation, also exchanging quantum information with the detector. With a gapless detector, the only ``rotation'' that takes place is due to the interaction with the quantum field. This is why, intuitively, gapless detectors can be better at extracting field correlations, as the only effect that they are sensitive to is the quantum field itself. This intuition is also aligned with the continuous variables studies of~\cite{MariaContinuous}.

Finally, notice that the lack of an energy gap does not prevent one from defining excited and ground states, as one can always consider that a magnetic field is applied before and after the interaction, but not while it is taking place. For this reason, we will keep our notation $\{\ket{g},\ket{e}\}$ for the basis of the qubit system, and we will keep the nomenclature of ``ground'' and ``excited'' states for the eigenvectors of $ \hat{\sigma}^+\hat{\sigma}^-$, even when $\Omega = 0$.

\subsection{A single gapless UDW detector interacting with a scalar quantum field}\label{sub:1gapless}

We first consider the case where a single particle detector interacts with a field, with the assumption that $\Omega = 0$. In this case, the interaction Hamiltonian reduces to
\begin{equation}
    \hat{h}_I(\mf x) = \lambda \Lambda(\mf x) \hat{\mu} \,\hat{\phi}(\mf x),
\end{equation}
and here we will consider that $\hat{\mu}$ is any constant operator in the qubit's Hilbert space. The fact that $\hat{\mu}$ is time independent implies an important technical fact: we will now have the microcausality condition fulfilled. Indeed,
\begin{equation}
    [\hat{h}_I(\mf x), \hat{h}_I(\mf x')] = \lambda^2 \Lambda(\mf x)\Lambda(\mf x') \hat{\mu}^2[\hat{\phi}(\mf x), \hat{\phi}(\mf x')],
\end{equation}
which will always commute whenever the points $\mf x$ and $\mf x'$ are spacelike separated. This fact prevents the incompatibilities with relativity described in~\cite{eduardoOld,us2,PipoFTL,mariaPipoNew} to take place. 

A consequence of the simple expression for the commutator $[\hat{h}_I(\mf x), \hat{h}_I(\mf x')]$ is that the time evolution operator for the detector and field can be solved non-perturbatively using the Magnus expansion. In essence, due to the fact that $[[\hat{h}_I(\mf x), \hat{h}_I(\mf x')],\hat{h}_I(\mf x'')] = 0$, we have that~\cite{magnus}
\begin{equation}
    \hat{U}_I = \mathcal{T}\exp\left(- \ii \int \dd V \hat{h}_I(\mf x)\right) = e^{\hat{\Theta}_1+\hat{\Theta}_2},
\end{equation}
where
\begin{align}
    \hat{\Theta}_1 &= - \ii \int \dd V \hat{h}_I(\mf x) = - \ii \lambda \hat{\mu} \hat{\phi}(\Lambda),\\
    \hat{\Theta}_2 &= - \frac{1}{2} \int \dd V \dd V' \theta(t-t')[\hat{h}_I(\mf x), \hat{h}_I(\mf x')] \\&= -  \ii \frac{\lambda^2}{2} \hat{\mu}^2G_R(\Lambda,\Lambda),
\end{align}
where we used that $[\hat{\phi}(\mf x), \hat{\phi}(\mf x')] = \ii E(\mf x,\mf x')$ and that $\theta(t-t')E(\mf x, \mf x') = G_R(\mf x,\mf x')$. We can then write the time evolution operator for the detector-field system as
\begin{equation}
    \hat{U}_I = e^{- \ii \lambda \hat{\mu} \hat{\phi}(\Lambda)}e^{- \ii \hat{\mu}^2 \mathcal{G}},
\end{equation}
where we used that $[\hat{\mu},\hat{\mu}^2] = 0$ in order to separate the exponentials and we denoted $\mathcal{G} = \frac{\lambda^2}{2} G_R(\Lambda,\Lambda)$ 

We will again assume that the detector and field start in an uncorrelated state $\hat{\rho}_0 = \hat{\rho}_{\tc{d},0}\otimes \hat{\rho}_\omega$, where $\hat{\rho}_\omega$ is a representation of a quasifree state $\omega$ for the quantum field. In this case, one can compute the final state of the detector by tracing over the field's state,
\begin{align}
    \hat{\rho}_\tc{d} &= \tr_\phi\left(\hat{U}_I\hat{\rho}_0\hat{U}_I^\dagger\right) \\&= e^{- \ii \hat{\mu}^2 \mathcal{G}}\tr_\phi\left( e^{- \ii \lambda \hat{\mu} \hat{\phi}(\Lambda)}(\hat{\rho}_{\tc{d},0}\otimes \hat{\rho}_\omega )e^{\ii \lambda \hat{\mu} \hat{\phi}(\Lambda)}\right)e^{\ii \hat{\mu}^2 \mathcal{G}}\nonumber,
\end{align}
with $\mathcal{G} = \frac{\lambda^2}{2} G_R(\Lambda,\Lambda)$. We proceed with the computation assuming $\hat{\mu}^2 = \openone$, and using the identities 
\begin{align}
    e^{-\ii \lambda \hat{\mu} \hat{\phi}(\Lambda)} &= \text{cos} 
 (\lambda\hat{\phi}(\Lambda)) - \ii \hat{\mu} \,\text{sin}(\lambda \hat{\phi}(\Lambda)),\nonumber\\
    \omega(e^{\ii \hat{\phi}(f)}) &= \omega(\text{cos}(\hat{\phi}(f)) = e^{-\frac{1}{2}W(f,f)},\nonumber\\
    e^{\ii \hat{\phi}(f)}e^{\ii \hat{\phi}(g)} &= e^{\ii \hat{\phi}(f+g)}e^{\frac{\ii}{2}E(f,g)},\nonumber\\
    \omega(\text{cos}^2(\hat{\phi}(f)) &= e^{-W(f,f)}\cosh(W(f,f)),\label{eq:identities}
\end{align}
so that we find
\begin{align}
    \hat{\rho}_\tc{d} =& \omega(\text{cos}^2(\lambda \hat{\phi}(\Lambda)) e^{- \ii \hat{\mu}^2 \mathcal{G}} \hat{\rho}_{\tc{d},0} e^{\ii\hat{\mu}^2 \mathcal{G}}\\
    &+ \omega(\text{sin}^2(\lambda \hat{\phi}(\Lambda)) e^{- \ii \mathcal{G}} \hat{\mu}\,\hat{\rho}_{\tc{d},0}\,\hat{\mu}e^{\ii \mathcal{G}}\nonumber\\
    =& \left(e^{-\xi}\cosh(\xi)\hat{\rho}_{\tc{d},0} + e^{-\xi}\sinh(\xi)\hat{\mu}\,\hat{\rho}_{\tc{d},0}\,\hat{\mu}\right),\nonumber
\end{align}
which establishes a quantum channel acting in the qubit with $\xi = \lambda^2 W(\Lambda,\Lambda)$. In the case of $\hat{\mu}= \hat{\sigma}^++\hat{\sigma}^-$, this quantum channel is a bit-flip channel with parameter \mbox{$p = e^{-\xi}\sinh(\xi)$}.

One can also use the expressions found in Section~\ref{sec:closed} to compute the parameter $\xi$ in the case where the detector is interacting with the Minkowski vacuum of a massless scalar field, with the spacetime smearing function
\begin{equation}
    \Lambda(\mf x) = e^{- \frac{t^2}{2T^2}}\frac{e^{- \frac{|\bm x|^2}{2\sigma^2}}}{(2\pi \sigma^2)^\frac{3}{2}}.
\end{equation}
The value for $\xi$ is then obtained from Eq.~\eqref{eq:Wpm} by taking $|\bm L| = t_0 = \Omega = 0$. We obtain:
\begin{align}
    W(\Lambda,\Lambda) &= \frac{1}{4\pi \alpha^2},
\end{align}
where again $\alpha = \sqrt{1+\sigma^2/T^2}$. Notice, in particular, that the purity of the state $\hat{\rho}_\tc{d}$ is entirely determined by the parameter $\xi$. Indeed, if  $\hat{\rho}_{\tc{d},0}$ starts in a pure state, we find
\begin{equation}
    \tr(\hat{\rho}_\tc{d}^2) = e^{-2\xi}(\cosh(2\xi) + M^2\sinh(2 \xi)),
\end{equation}
 with $M^2 = \tr((\hat{\mu}\,\hat{\rho}_{\tc{d,0}})^2)$. The qubit's purity is then a decreasing function of $\xi$. Using that $\xi$ increases with $T$, we can see that the purity of the state decreases with $T$, and asymptotically reaches its minimum value when $\xi \to \lambda^2/4\pi$, unless \mbox{$[\hat{\rho}_{\tc{d},0},\hat{\mu}] = 0$}, in which case the evolution is always unitary for the detector, as it starts in an eigenstate of the Hamiltonian. We also see that the stronger the coupling, the more mixed the detector state is, as $\xi \propto \lambda^2$. The fact that the detector in general ends in a mixed state shows that the detector generally becomes entangled with the field. 

\subsection{Two gapless detectors interacting with a scalar quantum field}\label{sub:2gapless}

A similar method to that of Subsection~\ref{sub:1gapless} can also be applied to the case where two UDW detectors interact with a quantum field. This case has been studied in~\cite{Landulfo} under the assumption that the interaction of one of the detectors happens before the other one. We will not make this assumption here, and will instead obtain results for arbitrary interaction regions for the two detectors.

In this case, the total interaction Hamiltonian density is given by
\begin{equation}
    \hat{h}_I(\mf x) = \lambda \hat{\mu}_\tc{a} \Lambda_\tc{a}(\mf x) \hat{\phi}(\mf x) + \lambda \hat{\mu}_\tc{b} \Lambda_\tc{b}(\mf x) \hat{\phi}(\mf x),
\end{equation}
and the Magnus expansion can again be used to compute the time evolution operator. We find
\begin{equation}
    \hat{U}_I = e^{\hat{\Theta}_1+\hat{\Theta}_2},
\end{equation}
where
\begin{align}
    \hat{\Theta}_1 &= - \ii \int \dd V \hat{h}_I(\mf x) = - \ii \lambda \hat{\mu}_\tc{a} \hat{\phi}(\Lambda_\tc{a})- \ii \lambda \hat{\mu}_\tc{b} \hat{\phi}(\Lambda_\tc{b}),\label{eq:TH2}\\
    \hat{\Theta}_2 &= - \frac{1}{2} \int \dd V \dd V' \theta(t-t')[\hat{h}_I(\mf x), \hat{h}_I(\mf x')] \\&= - \frac{\lambda^2}{2}\int \dd V \dd V' \theta(t-t')[\hat{\phi}(\mf x), \hat{\phi}(\mf x')]\nonumber\\
    &\:\:\:\:\:\:\:\:\:\:\:\:\:\:\:\:\:\:\:\:\:\:\:\:\Big(\hat{\mu}_\tc{a}^2 \Lambda_\tc{a}(\mf x)\Lambda_\tc{a}(\mf x')+\hat{\mu}_\tc{a}\hat{\mu}_\tc{b} \Lambda_\tc{a}(\mf x)\Lambda_\tc{b}(\mf x')\nonumber\\
    &\:\:\:\:\:\:\:\:\:\:\:\:\:\:\:\:\:\:\:\:\:\:\:\:+\hat{\mu}_\tc{b}\hat{\mu}_\tc{a} \Lambda_\tc{b}(\mf x)\Lambda_\tc{a}(\mf x')+\hat{\mu}_\tc{b}^2 \Lambda_\tc{b}(\mf x)\Lambda_\tc{b}(\mf x')\Big)\nonumber\\
    &= - \frac{\ii \lambda^2}{2}\Big(\hat{\mu}_\tc{a}^2 G_R(\Lambda_\tc{a},\Lambda_\tc{a})+\hat{\mu}_\tc{b}^2 G_R(\Lambda_\tc{b},\Lambda_\tc{b})\\
    &\:\:\:\:\:\:\:\:\:\:\:\:\:\:\:\:\:\:\:\:\:\:+\hat{\mu}_\tc{a}\hat{\mu}_\tc{b} \left(G_R(\Lambda_\tc{a},\Lambda_\tc{b})+ G_R(\Lambda_\tc{b},\Lambda_\tc{a})\right)\Big),\nonumber
\end{align}
where we used that $\theta(t-t')[\hat{\phi}(\mf x),\hat{\phi}(\mf x')] = \ii G_R(\mf x, \mf x')$, and that the operators $\hat{\mu}_\tc{a}$ and $\hat{\mu}_\tc{b}$ commute. We then denote $\Delta_\tc{ab} = \lambda^2(G_R(\Lambda_\tc{a},\Lambda_\tc{b})+G_A(\Lambda_\tc{a},\Lambda_\tc{b}))$, $\mathcal{G}_\tc{a} = \frac{\lambda^2}{2}G_R(\Lambda_\tc{a},\Lambda_\tc{a})$, and $\mathcal{G}_\tc{b} = \frac{\lambda^2}{2}G_R(\Lambda_\tc{b},\Lambda_\tc{b})$, so that Eq.~\eqref{eq:TH2} allows us to write
\begin{equation}
    \hat{\Theta}_2 = - \ii \hat{\mu}_\tc{a}^2\mathcal{G}_\tc{a}- \ii\hat{\mu}_\tc{b}^2 \mathcal{G}_\tc{b} -  \tfrac{\ii}{2}\hat{\mu}_\tc{a}\hat{\mu}_\tc{b}\Delta_\tc{ab}.
\end{equation}
The fact that the commutator $[\hat{h}_I(\mf x), \hat{h}_I(\mf x')]$ commutes with $h_I(\mf x'')$ implies that only $\hat{\Theta}_1$ and $\hat{\Theta}_2$ are non-zero in the Magnus expansion so that the unitary time evolution operator reads
\begin{equation}\label{eq:UI2g}
    \hat{U}_I = e^{- \ii \lambda\hat{\mu}_\tc{a}\hat{\phi}(\Lambda_\tc{a}) - \ii \lambda\hat{\mu}_\tc{b} \hat{\phi}(\Lambda_\tc{b})} e^{- \ii \hat{\mu}_\tc{a}^2\mathcal{G}_\tc{a}}e^{- \ii\hat{\mu}_\tc{b}^2 \mathcal{G}_\tc{b}}e^{-  \tfrac{\ii}{2}\hat{\mu}_\tc{a}\hat{\mu}_\tc{b}\Delta_\tc{ab}},
\end{equation}
where we used that $[\hat{\Theta}_1,\hat{\Theta}_2] = 0$ in order to separate the exponentials. One can also use the Baker-Campbell-Hausdorff formula in order to factor $\hat{U}_I$ as
\begin{align}
    \hat{U}_I &= e^{- \ii \lambda\hat{\mu}_\tc{a}\hat{\phi}(\Lambda_\tc{a})}e^{- \ii \lambda\hat{\mu}_\tc{b} \hat{\phi}(\Lambda_\tc{b})} e^{- \ii \hat{\mu}_\tc{a}^2\mathcal{G}_\tc{a}- \ii\hat{\mu}_\tc{b}^2 \mathcal{G}_\tc{b}}e^{-  \tfrac{\ii}{2}\hat{\mu}_\tc{a}\hat{\mu}_\tc{b}(\Delta_\tc{ab}-E_{\tc{ab}})}\nonumber\\
    &= e^{- \ii \lambda\hat{\mu}_\tc{b} \hat{\phi}(\Lambda_\tc{b})} e^{- \ii \lambda\hat{\mu}_\tc{a}\hat{\phi}(\Lambda_\tc{a})}e^{- \ii \hat{\mu}_\tc{a}^2\mathcal{G}_\tc{a}- \ii\hat{\mu}_\tc{b}^2 \mathcal{G}_\tc{b}}e^{-  \tfrac{\ii}{2}\hat{\mu}_\tc{a}\hat{\mu}_\tc{b}(\Delta_\tc{ab}+E_{\tc{ab}})},\label{eq:BCH}
\end{align}
where $E_{\tc{ab}} = \lambda^2 E(\Lambda_\tc{a},\Lambda_\tc{b})$, and we have
\begin{align}
    \tfrac{1}{2}(\Delta_{\tc{ab}} - E_\tc{ab}) &= \lambda^2G_A(\Lambda_\tc{a},\Lambda_\tc{b}) = \lambda^2G_R(\Lambda_\tc{b},\Lambda_\tc{a}),\\
    \tfrac{1}{2}(\Delta_{\tc{ab}} + E_\tc{ab}) &= \lambda^2G_R(\Lambda_\tc{a},\Lambda_\tc{b}) = \lambda^2G_A(\Lambda_\tc{b},\Lambda_\tc{a}).
\end{align}



In order to compute the final state of the two detectors after tracing the field it is more convenient to work with the expression from Eq. \eqref{eq:UI2g}. Assume that the initial state of the detectors-field systems is $\hat{\rho}_0 = \hat{\rho}_{\tc{ab},0}\otimes \hat{\rho}_\omega$, where, as usual, $\hat{\rho}_\omega$ is the representation of a quasifree state $\omega$ in the quantum field theory. We further assume that $\hat{\mu}_\tc{a}^2 = \hat{\mu}_\tc{b}^2 = \openone$, so that the effect of the local unitaries $e^{- \ii \hat{\mu}_\tc{a}^2\mathcal{G}_\tc{a}}e^{- \ii\hat{\mu}_\tc{b}^2 \mathcal{G}_\tc{b}}$ becomes negligible. Given that the unitary $e^{-  \tfrac{\ii}{2}\hat{\mu}_\tc{a}\hat{\mu}_\tc{b}\Delta_\tc{ab}}$ commutes with the field-dependent term, we can separate the action of the unitary
\begin{equation}
    \hat{U}_\phi = e^{- \ii \lambda\hat{\mu}_\tc{a}\hat{\phi}(\Lambda_\tc{a}) - \ii \lambda\hat{\mu}_\tc{b} \hat{\phi}(\Lambda_\tc{b})}
\end{equation}
from the rest. In order to proceed with the computations, we denote the eigenstate of $\hat{\mu}_\tc{a}$ and $\hat{\mu}_\tc{b}$ by $\ket{\pm_\tc{a}}$ and $\ket{\pm_\tc{b}}$, so that
\begin{align}
    \hat{U}_\phi \hat{\rho}_0\hat{U}_\phi^\dagger &= e^{- \ii \lambda\hat{\mu}_\tc{a}\hat{\phi}(\Lambda_\tc{a}) - \ii \lambda\hat{\mu}_\tc{b} \hat{\phi}(\Lambda_\tc{b})}\hat{\rho}_{0}e^{\ii \lambda\hat{\mu}_\tc{a}\hat{\phi}(\Lambda_\tc{a}) + \ii \lambda\hat{\mu}_\tc{b} \hat{\phi}(\Lambda_\tc{b})}\nonumber\\
    &=\!\!\!\!\sum_{{\substack{{}_{\mu_\tc{a},\mu_\tc{b} = \pm}\\{}_{\mu_\tc{a}',\mu_\tc{b}' = \pm}}}} \!\!e^{- \ii \lambda\hat{\phi}({\mu}_\tc{a}\Lambda_\tc{a}+{\mu}_\tc{b}\Lambda_\tc{b})}\hat{\rho}_\omega e^{\ii \lambda\hat{\phi}({\mu}_\tc{a}'\Lambda_\tc{a}+{\mu}_\tc{b}'\Lambda_\tc{b})}\nonumber\\[-16pt]
    &\:\:\:\:\:\:\:\:\:\:\:\:\:\:\:\:\:\:\:\:\:\:\times\bra{\mu_\tc{a}\mu_{\tc{b}}}\hat{\rho}_{\tc{ab},0}\ket{\mu_\tc{a}'\mu_{\tc{b}'}}  \ket{\mu_\tc{a}\mu_{\tc{b}}}\!\!\bra{\mu_\tc{a}'\mu_{\tc{b}'}}.
\end{align}
The next step is to trace over the field to obtain the state $\hat{\sigma}_\tc{ab} = \tr_\phi(\hat{U}_\phi \hat{\rho}_0\hat{U}_\phi^\dagger)$. We find
\begin{align}
    &\hat{\sigma}_\tc{ab} 
    =\!\!\!\!\sum_{{\substack{{}_{\mu_\tc{a},\mu_\tc{b} = \pm}\\{}_{\mu_\tc{a}',\mu_\tc{b}' = \pm}}}} \!\!\omega\!\left(e^{\ii \lambda\hat{\phi}({\mu}_\tc{a}'\Lambda_\tc{a}+{\mu}_\tc{b}'\Lambda_\tc{b})}e^{- \ii \lambda\hat{\phi}({\mu}_\tc{a}\Lambda_\tc{a}+{\mu}_\tc{b}\Lambda_\tc{b})}\right)\nonumber\\[-15pt]
    &\:\:\:\:\:\:\:\:\:\:\:\:\:\:\:\:\:\:\:\:\:\:\:\:\:\:\:\:\:\:\:\:\:\:\:\:\:\:\:\times\bra{\mu_\tc{a}\mu_{\tc{b}}}\hat{\rho}_{\tc{ab},0}\ket{\mu_\tc{a}'\mu_{\tc{b}'}}  \ket{\mu_\tc{a}\mu_{\tc{b}}}\!\!\bra{\mu_\tc{a}'\mu_{\tc{b}'}}\nonumber\\[10pt]
    &=\!\!\!\!\sum_{{\substack{{}_{\mu_\tc{a},\mu_\tc{b} = \pm}\\{}_{\mu_\tc{a}',\mu_\tc{b}' = \pm}}}} \!\!\! e^{\frac{\ii\lambda^2}{2}E(\mu_\tc{a}'\Lambda_\tc{a} + \mu_{\tc{b}}'\Lambda_\tc{b},\mu_\tc{a}\Lambda_\tc{a} + \mu_{\tc{b}}\Lambda_\tc{b})- \frac{\lambda^2}{2}||(\mu_{\tc{a}} - \mu_{\tc{a}}')\Lambda_{\tc{a}} + (\mu_{\tc{b}} - \mu_{\tc{b}}')\Lambda_{\tc{b}}||^2}\nonumber\\[-16pt]
    &\:\:\:\:\:\:\:\:\:\:\:\:\:\:\:\:\:\:\:\:\:\:\:\:\:\:\:\:\:\:\:\:\:\:\:\:\:\:\:\times\bra{\mu_\tc{a}\mu_{\tc{b}}}\hat{\rho}_{\tc{ab},0}\ket{\mu_\tc{a}'\mu_{\tc{b}'}}  \ket{\mu_\tc{a}\mu_{\tc{b}}}\!\!\bra{\mu_\tc{a}'\mu_{\tc{b}'}},
\end{align}
where we used
\begin{equation}
    \omega\left(e^{\ii \lambda \hat{\phi}(f)}e^{\ii \lambda \hat{\phi}(g)}\right) = e^{-\frac{\ii \lambda^2}{2}E(f,g) - \frac{\lambda^2}{2} W(f+g,f+g)},
\end{equation}
and we denoted $||f||^2 = W(f,f)$. We can now incorporate the unitary $e^{-  \tfrac{\ii}{2}\hat{\mu}_\tc{a}\hat{\mu}_\tc{b}\Delta_\tc{ab}}$ again, so that the final state of the detectors is given by
\begin{equation}
    \hat{\rho}_\tc{ab} = e^{-  \tfrac{\ii}{2}\hat{\mu}_\tc{a}\hat{\mu}_\tc{b}\Delta_\tc{ab}} \hat{\sigma}_\tc{ab}e^{\tfrac{\ii}{2}\hat{\mu}_\tc{a}\hat{\mu}_\tc{b}\Delta_\tc{ab}}.\label{eq:sab}
\end{equation} 
Once again, notice that the final state of the qubits is entirely given in terms of bi-distributions of the quantum field smeared against $\Lambda_\tc{a}(\mf x)$ and $\Lambda_{\tc b}(\mf x')$. 

Finally, we write $\rho_{ij}$, $i,j = 1,...,4$ for the components of $\hat{\rho}_{\tc{ab},0}$ in the basis $\{\ket{+_\tc{a}+_\tc{b}},\ket{+_\tc{a}-_\tc{b}},\ket{-_\tc{a}+_\tc{b}},\ket{-_\tc{a}-_\tc{b}}\}$, so that the final state of the detectors state can be written (in this same basis) as
\begin{widetext}
\begin{equation}
    \hat{\rho}_\tc{ab} = \begin{pmatrix}
        \rho_{11} & e^{-2W_{\tc{bb}} + \ii(E_\tc{ab} - \Delta_{\tc{ab}})}\rho_{12} & e^{-2W_{\tc{aa}} - \ii(E_\tc{ab} + \Delta_{\tc{ab}})}\rho_{13} & e^{-2(W_{\tc{aa}}+W_{\tc{bb}}+H_{\tc{ab}})}\rho_{14} \\
        e^{-2W_{\tc{bb}} - \ii(E_\tc{ab} - \Delta_{\tc{ab}})}\rho_{21} & \rho_{22} & e^{-2(W_{\tc{aa}}+W_{\tc{bb}}-H_{\tc{ab}})}\rho_{23} & e^{-2W_{\tc{aa}} + \ii(E_\tc{ab} + \Delta_{\tc{ab}})}\rho_{24} \\
        e^{-2W_{\tc{aa}} + \ii(E_\tc{ab} + \Delta_{\tc{ab}})}\rho_{31} & e^{-2(W_{\tc{aa}}+W_{\tc{bb}}-H_{\tc{ab}})}\rho_{32} & \rho_{33} & e^{-2W_{\tc{bb}} - \ii(E_\tc{ab} - \Delta_{\tc{ab}})}\rho_{34} \\
        e^{-2(W_{\tc{aa}}+W_{\tc{bb}}+H_{\tc{ab}})}\rho_{41} & e^{-2W_{\tc{aa}} - \ii(E_\tc{ab} + \Delta_{\tc{ab}})}\rho_{42} & e^{-2W_{\tc{bb}} + \ii(E_\tc{ab} - \Delta_{\tc{ab}})}\rho_{43} & \rho_{44} 
    \end{pmatrix}.
\end{equation}
\end{widetext}
For each of the bi-distributions $W, H, E$, and $\Delta$, we use the convention $B_{\tc{ab}} = \lambda^2 B(\Lambda_\tc{a},\Lambda_\tc{b})$.

\begin{figure}[h!]
    \centering
    \includegraphics[width=8.6cm]{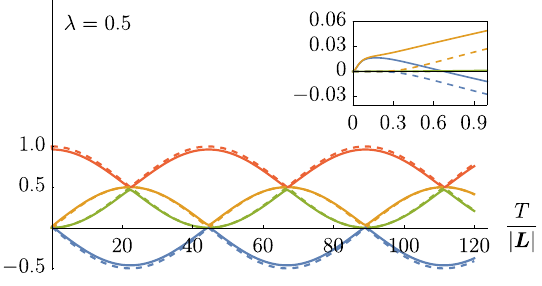}
    \includegraphics[width=8.6cm]{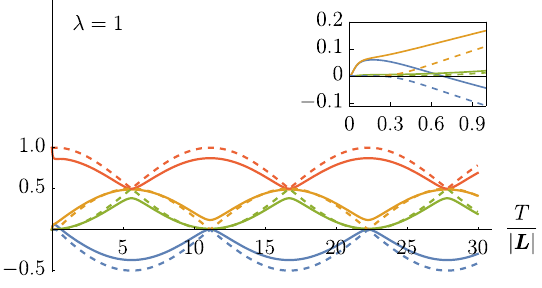}
    \includegraphics[width=8.6cm]{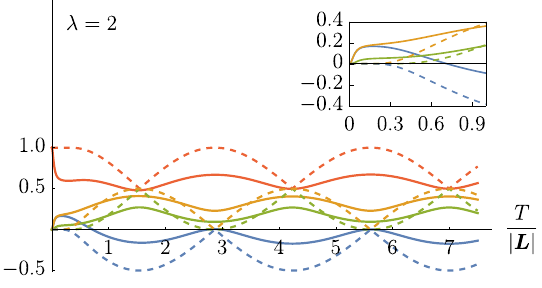}
    \caption{The solid lines correspond to the eigenvalues of the partial transpose of the detectors final state $\hat{\rho}_\tc{ab}^{t_\tc{b}}$, when the detectors start both in their ground state, as a function of the interaction time $T$, scaled by the detectors' separation $|\bm L|$. The dashed lines are the eigenvalues of $\hat{\rho}_\tc{ab}^{t_\tc{b}}$ if it were to evolve only according to the unitary \mbox{$\hat{U}_\tc{c} = e^{- \frac{\ii}{2}\hat{\mu}_\tc{a} \hat{\mu}_\tc{b} \Delta_\tc{ab}}$}. We picked  $\sigma = 0.05|\bm L|$ for these plots.}
    \label{fig:t}
\end{figure}

In order to see an explicit example, we consider the case where $\hat{\mu}_\tc{a} = \hat{\sigma}_\tc{a}^+ + \hat{\sigma}_\tc{a}^-$, $\hat{\mu}_\tc{b} = \hat{\sigma}_\tc{b}^+ + \hat{\sigma}_\tc{b}^-$, the detectors undergo inertial trajectories in Minkowski spacetime, and interact with the vacuum of a massless scalar field in Gaussian spacetime regions. For convenience, we use the same spacetime smearing functions as in Eqs.~\eqref{eq:LA} and~\eqref{eq:LB} with $t_0 = 0$. We will assume both detectors to start in their ground states, with \mbox{$\hat{\rho}_{\tc{ab},0} = \ket{g_\tc{a}}\!\!\bra{g_\tc{a}}\otimes \ket{g_\tc{b}}\!\!\bra{g_\tc{b}}$}. As we did in the example of entanglement harvesting, we will be interested in checking the conditions so that the detectors can end up in an entangled state. In order to check that, we plot the eigenvalues of the partial transpose of $\hat{\rho}_{\tc{ab}}$ in Fig.~\ref{fig:t} as a function of the effective interaction time $T$ for three different values of the coupling constant. The dashed lines represent the eigenvalues of $\hat{\rho}_{\tc{c}}^{t_\tc{b}}$, where we define
\begin{equation}
    \hat{\rho}_\tc{c} = \hat{U}_\tc{c} \hat{\rho}_{\tc{ab},0} \hat{U}_\tc{c}^\dagger, \quad\quad \hat{U}_\tc{c} = e^{- \frac{\ii}{2} \hat{\mu}_\tc{a}\hat{\mu}_\tc{b}\Delta_{\tc{ab}}}.
\end{equation}
We show the eigenvalues of $\hat{\rho}_\textsc{c}$ for comparison. 

Keep in mind that the detectors are entangled if and only if the partial transpose of their density operator has a negative eigenvalue. Notice that in this example, for small values of the coupling constant $\lambda$, the behaviour of the state evolved by the unitary $\hat{U}_\tc{c} = e^{- \frac{\ii}{2} \hat{\mu}_\tc{a}\hat{\mu}_\tc{b}\Delta_{\tc{ab}}}$ is very similar to the interaction with the quantum field. This is because for small coupling constants, the detectors are not subject to too much noise, and they are able to communicate through the field without getting too entangled with the field itself. 

\vspace{-1mm}

Also notice that for small values of $T$, the detectors cannot become entangled. This can be seen in the extended plots on the top right, which display the region $\frac{T/|\bm L|} < 1$. The fact that the detectors cannot become entangled when $T\lesssim |\bm L|$ is in agreement with the results of the no-go theorems proven in~\cite{HarvestingQueNemLouko,nogo}, where it was proven that gapless detectors cannot harvest if their interaction regions are spacelike separated. The fact that the detectors start becoming entangled for $T \approx 0.7 |\bm L| <|\bm L|$ is merely an artifact of the non-compact support of the Gaussian switching functions considered, so that for $T\gtrsim0.7|\bm L|$, there is still enough causal contact for the detectors to become entangled. Importantly, regardless of the coupling constant, we find that the detectors are only able to become entangled for $T \gtrsim 0.7 |\bm L|$, showing that the detectors' inability to communicate is independent of how strongly they couple to the field.

\begin{figure}[h!]
    \centering
    \includegraphics[width=8.6cm]{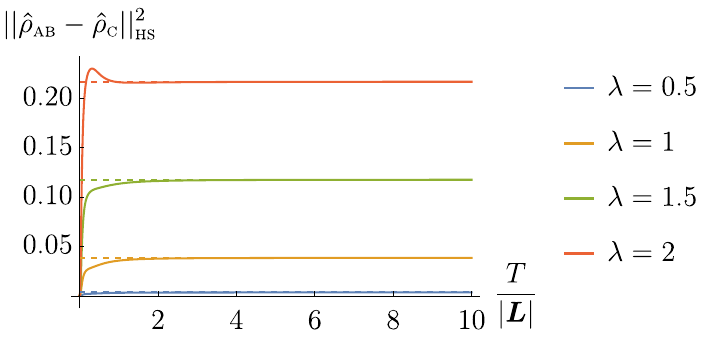}
    \caption{The squared Hilbert-Schmidt norm of the difference between the density operators $\hat{\rho}_\tc{ab}$ and $\hat{\rho}_\tc{c}$ for $\sigma = 0.05 |\bm L|$ considering detectors that start in their ground state. The dashed lines correspond to their asymptotic limit as $T\to\infty$.}
    \label{fig:pass}
\end{figure}

We end our analysis of this specific example by computing the explicit difference between the final density operator $\hat{\rho}_\tc{ab}$, obtained considering the complete interaction with the field, and the operator $\hat{\rho}_\tc{c}$, obtained by considering only the unitary part of the evolution, $\hat{U}_\tc{c}$. In Fig.~\ref{fig:pass}, we plot the squared Hilbert-Schmidt norm\footnote{The Hilbert-Schmidt norm is a norm in the space of operators in a finite dimensional vector space. It is defined as $||\hat{A}||_\tc{hs} = \sqrt{\Tr(A^\dagger A)}$.} of the difference $\hat{\rho}_\tc{ab} - \hat{\rho}_\tc{c}$ for different values of $\lambda$, using the same setup as we considered before. In the figure we see that the norm of the difference quickly goes to zero as $\lambda$ decreases. We can also notice that in the limit of $T\to\infty$, the difference between the two evolutions becomes a constant.

\begin{figure}[h!]
    \centering
    \includegraphics[width=8.6cm]{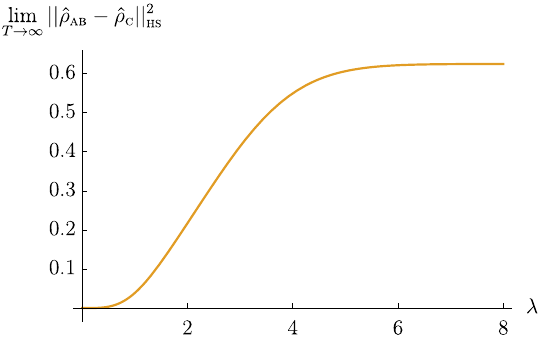}
    \caption{The limit of the asymptotic behaviour of the squared Hilbert-Schmidt norm of the operator $\rho_{\tc{ab}} - \hat{\rho}_\tc{c}$. This limit only depends on $\lambda$.}
    \label{fig:passLimit}
\end{figure}

Given that we have access to analytical expressions for both $\hat{\rho}_\tc{ab}$ and $\hat{\rho}_{\tc{c}}$, we can also compute the asymptotic behaviour of $||\hat{\rho}_\tc{ab} - \hat{\rho}_\tc{c}||^2_\tc{hs}$ in the limit where $\sigma\ll |\bm L|$ and $T\gg |\bm L|$ (with the assumption of $\sigma T \ll |\bm L|^2$). In this limit, we find that the result is independent of $\sigma$, $|\bm L|$, and, of course, $T$. We find
\begin{align}
    \lim_{T\to\infty}||\hat{\rho}_\tc{ab} & - \hat{\rho}_\tc{c}||^2_\tc{hs} \\
    &\!\!= \frac{1}{8}\left(5 + e^{-\frac{4 \lambda^2}{\pi}}-2e^{-\frac{2 \lambda^2}{\pi}}+4e^{-\frac{\lambda^2}{\pi}} - 8 e^{-\frac{ \lambda^2}{2\pi}}\right).\nonumber
\end{align}
We plot this result in Fig.~\ref{fig:passLimit}. Also notice that the behaviour of the above limit for $\lambda\to 0$ is given by
\begin{align}
   \lim_{T\to\infty}||\hat{\rho}_\tc{ab} & - \hat{\rho}_\tc{c}||^2_\tc{hs} = \frac{5 \lambda^4}{8 \pi^2} + \mathcal{O}(\lambda^6),
\end{align}
suggesting that indeed, in the regime of small coupling constant, the evolution is well modelled by the unitary $\hat{U}_\tc{c}$.

\subsection{Recovering previous results in the literature}\label{sub:comparison}

In this subsection we will recover the results of~\cite{Landulfo} in the case where the interaction of $\tc{A}$ happens ``before'' the interaction of $\tc{B}$. In this context, the word ``before'' means that there exists a Cauchy surface $\Sigma$ separating the supports of $\Lambda_\tc{a}$ and $\Lambda_\tc{b}$ such that $\Lambda_\tc{a}$ is non-zero only in the causal past of $\Sigma$ and $\Lambda_\tc{b}$ is non-zero only in the causal future of $\Sigma$. In order to match our results more clearly, we will also assume that $\hat{\mu}_\tc{a}^2 = \hat{\mu}_\tc{b}^2 = \openone$, so that one can get rid of the local unitaries, as they contribute only with global phases, and do not affect the time evolution. The time ordering operation ensures that one can write the unitary time evolution operator as
\begin{equation}\label{eq:UIErickson}
   \hat{U}_I = e^{- \ii \lambda\hat{\mu}_\tc{b} \hat{\phi}(\Lambda_\tc{b})} e^{- \ii \lambda\hat{\mu}_\tc{a}\hat{\phi}(\Lambda_\tc{a})},
\end{equation}
with the unitary first being applied to the field and qubit $\tc{A}$, and then being applied to the field and qubit $\tc{B}$. However, in~\cite{Landulfo}, the interaction is written in a different from, with the unitary first being applied to $\tc{B}$ and only later to $\tc{A}$. Eq.~\eqref{eq:UIErickson} can be put in this form by commuting the two unitaries using the Baker-Campbell-Hausdorf formula, as was explicitly shown in Eq.~\eqref{eq:BCH}. Notice that if the interaction of $\tc{A}$ happens before $\tc{B}$, we have $\frac{1}{2}(\Delta_{\tc{ab}} + E_\tc{ab}) = \lambda^2G_R(\Lambda_\tc{a},\Lambda_\tc{b}) = 0$ and $\frac{1}{2}(\Delta_{\tc{ab}} - E_\tc{ab}) = \lambda^2G_R(\Lambda_\tc{b},\Lambda_\tc{a}) = - E_{\tc{ab}}$. Under these assumptions~\eqref{eq:BCH} reads
\begin{align}
    \hat{U}_I &= e^{- \ii \lambda\hat{\mu}_\tc{a}\hat{\phi}(\Lambda_\tc{a})}e^{- \ii \lambda\hat{\mu}_\tc{b} \hat{\phi}(\Lambda_\tc{b})} e^{\ii\hat{\mu}_\tc{a}\hat{\mu}_\tc{b}E_{\tc{ab}}}\nonumber\\
    &= e^{- \ii \lambda\hat{\mu}_\tc{b} \hat{\phi}(\Lambda_\tc{b})} e^{- \ii \lambda\hat{\mu}_\tc{a}\hat{\phi}(\Lambda_\tc{a})},\label{eq:UIcomp}
\end{align}
the first line of which is exactly Eq. (35) of reference~\cite{Landulfo}, with the appropriate changes of conventions and ignoring the irrelevant global phase terms.

We also notice that when the interaction of qubit $\tc{A}$ happens before the interaction of qubit $\tc{B}$ and they start uncorrelated, one can write the final state of $\tc{B}$ after the interaction as
\begin{align}
    \hat{\rho}_\tc{b} &= \tr_{\phi,\tc{a}}(\hat{U}_{\tc{b}\phi} \hat{U}_{\tc{a}\phi}(\hat{\rho}_{\tc{b},0} \otimes \hat{\rho}_{\tc{a},0} \otimes \hat{\rho}_\phi)\hat{U}_{\tc{a}\phi}^\dagger\hat{U}_{\tc{b}\phi}^\dagger)\\
    &= \tr_{\phi}(\hat{U}_{\tc{b}\phi} (\hat{\rho}_{\tc{b},0} \otimes \tr_\tc{a}(\hat{U}_{\tc{a}\phi} (\hat{\rho}_{\tc{a},0} \otimes  \hat{\rho}_\phi)\hat{U}_{\tc{a}\phi}^\dagger)\hat{U}_{\tc{b}\phi}^\dagger)\\
    &= \tr_{\phi}(\hat{U}_{\tc{b}\phi} (\hat{\rho}_\tc{b} \otimes \hat{\tilde{\rho}}_\phi)\hat{U}_{\tc{b}\phi}^\dagger),
\end{align}
where $\hat{\tilde{\rho}}_\phi$ is an updated state for the field after the interaction of qubit $\tc{A}$. As stated in~\cite{Landulfo}, as well as in~\cite{nogo}, this is an entanglement breaking channel for qubit $\tc{B}$, thus showing that $\tc{A}$ and $\tc{B}$ cannot become entangled if there is a Cauchy surface separating the support of their interaction regions.

Finally, we comment on the case of delta-coupled particle detectors (or ultra-fast switchings), in which case the supports of $\Lambda_\tc{a}$ and $\Lambda_\tc{b}$ are contained in spacelike surfaces. In this setup the energy gaps of the detectors become irrelevant, and the results of gapless detectors apply. Unless the supports of $\Lambda_\tc{a}$ and $\Lambda_\tc{b}$ overlap, it is always possible to find a Cauchy surface between the interaction regions, and the results discussed above also apply, so that Eq.~\eqref{eq:UIcomp} can be used to compute the final state of the detectors, and the qubits cannot end in an entangled state, confirming the results of~\cite{ericksonCapacity}, as well as another example of the no-go theorems of~\cite{HarvestingQueNemLouko,nogo}. Overall, we see that the results of this section generalize the previous results in the literature for gapless detectors as well as for delta-coupled particle detectors.

\section{Conclusions}\label{sec:conclusions}

We have studied the Wightman function, the Feynman propagator, the retarded and advanced Green's functions, the causal propagator and the symmetric propagator of a massless scalar field in the vacuum of Minkowski spacetime when smeared against Gaussian spacetime functions. We found closed-form expressions for each of these smeared bi-distributions, and analyzed examples where these correlations can be accessed by local probes that couple to the field in Gaussian-shaped regions. Our results allowed us to find easily evaluatable closed-form expressions for the final state of these probes in the regime of weak interactions with the field, and in the case where the probes have trivial internal dynamics during the interaction with the field.

In the case of weak interactions, we found closed-form expressions for the entanglement acquired by detectors that interact with a quantum field for a time $T$ and that are separated by a distance $L$. Using our results we found a simple expression for the contribution of signalling to the entanglement between the detectors. We also looked at asymptotic results of entanglement harvesting, and we found that for large spatial separations, the available entanglement in Gaussian spacetime regions behaves as $\ell^{-4}e^{- \ell^2/4}$, where $\ell = L/T$ is the distance between the regions relative to the interaction time. This confirms that the available entanglement between spacelike separated regions decreases exponentially with their distance. By comparing our results with the usual numerical integration methods used to study entanglement harvesting, we could also find the analytical expression for an integral that had not been solved in closed-form until this point (see Eq.~\eqref{eq:newInt}).

We also considered the case where the detectors that interact with the field have zero energy gap, generalizing the results of~\cite{Landulfo}. Using these general results, combined with the results of the bi-distributions, we studied explicit examples of gapless detectors interacting in Gaussian regions of spacetime, and found that gapless detectors can become entangled, provided that their interaction regions cannot be separated by a Cauchy surface. We also found that, for small coupling constants and spacetime regions separated by translations in space, the noise generated by the quantum field is suppressed, and the quantum gate established between the two detectors is well approximated by the unitary $e^{-  \tfrac{\ii}{2}\hat{\mu}_\tc{a}\hat{\mu}_\tc{b}\Delta_\tc{ab}}$, where $\Delta_\tc{ab}$ represents the symmetric exchange of information between the interaction regions. Finally, we discussed how these results are compatible with the no-go theorems of~\cite{HarvestingQueNemLouko,nogo} which prevent spacelike entanglement harvesting using gapless detectors. 

In Appendix~\ref{app}, we have also generalized the closed-form results presented in the manuscript to two-point functions between more general field observables (such as the field's momentum), and more general parameters of the Gaussian spacetime regions. We hope that the results and expressions presented in this manuscript will be useful for future studies of local operations in quantum field theory. The fact that all observables localized in Gaussian spacetime regions in a massless scalar QFT can be written in terms of the bi-distributions that have been computed in this manuscript showcases the wide range of applications of our results.\\

\vspace{6mm}

\begin{acknowledgements}

The author thanks Erickson Tjoa and Cisco Gooding for insightful discussions and the organizers of the ``Golden Wedding of Black Holes and Thermodynamics'' online conference, with special thanks to N\'ickolas Alves for the invitation to deliver a minicourse in Relativistic Quantum Information, which led the author to perform many of the calculations that resulted in this manuscript. The author also thanks David Kubiznak at Charles University, Robert Jonsson at the Nordic Institute of Theoretical Physics (NORDITA), \v{C}aslav Brukner at the Institute of Quantum Optics and Quantum Information (IQOQI), Dennis R\"atzel at the Center for Space Technology and Microgravity (ZARM) and Cisco Gooding at the University of Nottingham for hosting him during the production of this manuscript. The author also thanks his supervisors, Eduardo Mart\'in-Mart\'inez and David Kubiznak, for encouraging him to pursue this independent research project. TRP acknowledges support from the Natural Sciences and Engineering Research Council of Canada (NSERC) via the Vanier Canada Graduate Scholarship. Research at Perimeter Institute is supported in part by the Government of Canada through the Department of Innovation, Science and Industry Canada and by the Province of Ontario through the Ministry of Colleges and Universities. Perimeter Institute and the University of Waterloo are situated on the Haldimand Tract, land that was promised to the Haudenosaunee of the Six Nations of the Grand River, and is within the territory of the Neutral, Anishinaabe, and Haudenosaunee people.
\end{acknowledgements}

\onecolumngrid

\appendix

\section{More general parameters for the smeared bi-distributions}\label{app}

In this appendix we will compute the bi-distributions $W(f_1,f_2)$, $E(f_1,f_2)$ and $H(f_1,f_2)$ for general parameters  $(T_1,t_1,\Omega_1, \sigma_1, \bm L_1)$ and $(T_2,t_2,\Omega_2, \sigma_2, \bm L_2)$ for the functions $f_1(\mf x)$ and $f_2(\mf x)$ detailed in Section~\ref{sec:closed}. We will also compute $G_R(f_1,f_2)$, $G_A(f_1,f_2)$, $\Delta(f_1,f_2)$, and $G_F(f_1,f_2)$ in the case $\sigma_1 = \sigma_2 = \sigma$. Finally, we will discuss how to generalize our results for the computation of two-point functions that involve the momentum of a real massless quantum field and more general time derivatives of the field.

We start from Eq.~\eqref{eq:WABk}, and notice that for the functions
\begin{equation}
    F_\tc{i}(\bm x) = \frac{e^{-\frac{|\bm x - \bm L_\tc{i}|^2}{2 \sigma_\tc{i}^2}}}{(2\pi \sigma_\tc{i}^2)^{3/2}}, \quad \quad \chi_\tc{i}(t) = e^{- \frac{(t-t_\tc{i})^2}{2T_\tc{i}^2}}e^{\ii \Omega_\tc{i} t},
\end{equation}
we have the Fourier transforms
\begin{equation}
    \tilde{F}(\bm k) = e^{\ii \bm k \cdot \bm L_\tc{i} - \frac{\sigma_\tc{i}^2 |\bm k|^2}{2}}, \quad \quad \tilde{\chi}(\omega) = \sqrt{2 \pi} T_\tc{i} e^{\ii (\omega+\Omega_\tc{i}) t_\tc{i} - \frac{(\omega+\Omega_\tc{i})^2 T_\tc{i}^2}{2}}.
\end{equation}
Plugging the results above into Eq.~\eqref{eq:WABk}, we find
\begin{align}
    W(f_1,f_2) &= \frac{1}{(2\pi)^2} \int \frac{\dd^3\bm k}{2 |\bm k|} e^{\ii \bm k \cdot \bm L_1 - \frac{\sigma_1^2 |\bm k|^2}{2}} e^{-\ii \bm k \cdot \bm L_2 - \frac{\sigma_2^2 |\bm k|^2}{2}} T_2 e^{\ii (|\bm k|+\Omega_2) t_2 - \frac{(|\bm k|+\Omega_2)^2 T_2^2}{2}} T_1 e^{-\ii (|\bm k|-\Omega_1) t_1 - \frac{(|\bm k|-\Omega_1)^2 T_1^2}{2}}\nonumber\\
    &= \frac{T_1T_2e^{\ii (\Omega_1 t_1+\Omega_2 t_2)}}{(2\pi)^2} \int \frac{\dd^3\bm k}{2 |\bm k|} e^{\ii \bm k \cdot (\bm L_1 - \bm L_2)} e^{-\ii |\bm k|( t_1-t_2)}e^{ - \frac{(\sigma_1^2+\sigma_2^2) |\bm k|^2}{2}}   e^{- \frac{(|\bm k|+\Omega_2)^2 T_2^2}{2} - \frac{(|\bm k|-\Omega_1)^2 T_1^2}{2}} \nonumber\\
    &= \frac{T_1T_2e^{\ii (\Omega_1 t_1+\Omega_2 t_2)}}{(2\pi)^2} e^{- \frac{\Omega_2^2 T_2^2}{2} - \frac{\Omega_1^2 T_1^2}{2}}\int \frac{\dd^3\bm k}{2 |\bm k|} e^{\ii \bm k \cdot (\bm L_1 - \bm L_2)} e^{-\ii |\bm k|( t_1-t_2)}e^{ - \frac{(\sigma_1^2+\sigma_2^2+T_1^2 + T_2^2) |\bm k|^2}{2}}    e^{- |\bm k| (\Omega_2 T_2^2-\Omega_1 T_1^2)}\nonumber
    \\
    &= \frac{T_1T_2e^{\ii (\Omega_1 t_1+\Omega_2 t_2)}}{2\pi} e^{- \frac{\Omega_2^2 T_2^2}{2} - \frac{\Omega_1^2 T_1^2}{2}}\int \frac{\dd |\bm k|}{2 |\bm k|} |\bm k|^2 2 \,\text{sinc}(|\bm k|  |\bm L|) e^{-\ii |\bm k|t_0}e^{ - \frac{(\sigma_1^2+\sigma_2^2+T_1^2 + T_2^2) |\bm k|^2}{2}}    e^{- |\bm k| (\Omega_2 T_2^2-\Omega_1 T_1^2)}\nonumber\\
    &= \frac{T_1T_2e^{\ii (\Omega_1 t_1+\Omega_2 t_2)}}{2\pi|\bm L|} e^{- \frac{\Omega_2^2 T_2^2}{2} - \frac{\Omega_1^2 T_1^2}{2}}\int \dd |\bm k|\,\text{sin}(|\bm k|  |\bm L|) e^{-\ii |\bm k|t_0}e^{ - \frac{(\sigma_1^2+\sigma_2^2+T_1^2 + T_2^2) |\bm k|^2}{2}}    e^{- |\bm k| (\Omega_2 T_2^2-\Omega_1 T_1^2)},\label{eq:WfinalInt}
\end{align}
where $\bm L = \bm L_1 - \bm L_2$ and $t_0 = t_1-t_2$. Using the result
\begin{equation}
    \int_0^\infty \dd r e^{- \ii b r}e^{- \frac{a^2r^2}{2}} = \frac{\sqrt{\pi}}{\sqrt{2}a}e^{- \frac{b^2}{2a^2}}\Big(1 - \ii \text{erfi}\left(\frac{b}{\sqrt{2}a}\right)\Big),
\end{equation}
and writing $\sin(|\bm k||\bm L|)$ as exponentials we find
\begin{align}
     W(f_1,f_2) &= \frac{T_1T_2e^{\ii (\Omega_1 t_1+\Omega_2 t_2)}}{2\pi|\bm L|}  \frac{1}{2\ii}\frac{\sqrt{\pi}e^{- \frac{\Omega_2^2 T_2^2}{2} - \frac{\Omega_1^2 T_1^2}{2}}}{\sqrt{2} \sqrt{T_1^2 + T_2^2 + \sigma_1^2 + \sigma_2^2}}\Bigg(e^{- \frac{(t_0 - |\bm L|+\ii (\Omega_1 T_1^2- \Omega_2 T_2^2))^2}{2(T_1^2 + T_2^2 + \sigma_1^2 + \sigma_2^2)}}\Bigg(1 - \ii \text{erfi}\left(\tfrac{t_0 - |\bm L|+\ii (\Omega_1 T_1^2- \Omega_2 T_2^2)}{\sqrt{2}\sqrt{T_1^2 + T_2^2 + \sigma_1^2 + \sigma_2^2}}\right)\Bigg)\nonumber\\
     &\:\:\:\:\:\:\:\:\:\:\:\:\:\:\:\:\:\:\:\:\:\:\:\:\:\:\:\:\:\:\:\:\:\:\:\:\:\:\:\:\:\:\:\:\:\:\:\:\:\:\:\:\:\:\:\:\:\:\:\:\:\:\:\:\:\:\:\:\:\:\:\:\:\:\:\:\:\:\:\:\:\:\:\:\:\:\:\:\:\:\:\:\:-e^{- \frac{(t_0 + |\bm L|+\ii (\Omega_1 T_1^2- \Omega_2 T_2^2))^2}{2(T_1^2 + T_2^2 + \sigma_1^2 + \sigma_2^2)}}\Bigg(1 - \ii \text{erfi}\left(\tfrac{t_0 + |\bm L|+\ii (\Omega_1 T_1^2- \Omega_2 T_2^2)}{\sqrt{2}\sqrt{T_1^2 + T_2^2 + \sigma_1^2 + \sigma_2^2}}\right)\Bigg)\nonumber\\
     &= \frac{T_1T_2e^{\ii (\Omega_1 t_1+\Omega_2 t_2)}e^{- \frac{\Omega_2^2 T_2^2}{2} - \frac{\Omega_1^2 T_1^2}{2}}}{4\sqrt{2\pi}|\bm L|\sqrt{T_1^2 + T_2^2 + \sigma_1^2 + \sigma_2^2}} \Bigg(e^{- \frac{(t_0 - |\bm L|+\ii (\Omega_1 T_1^2- \Omega_2 T_2^2))^2}{2(T_1^2 + T_2^2 + \sigma_1^2 + \sigma_2^2)}}\Bigg(-\ii - \text{erfi}\left(\tfrac{t_0 - |\bm L|+\ii (\Omega_1 T_1^2- \Omega_2 T_2^2)}{\sqrt{2}\sqrt{T_1^2 + T_2^2 + \sigma_1^2 + \sigma_2^2}}\right)\Bigg)\nonumber\\
     &\:\:\:\:\:\:\:\:\:\:\:\:\:\:\:\:\:\:\:\:\:\:\:\:\:\:\:\:\:\:\:\:\:\:\:\:\:\:\:\:\:\:\:\:\:\:\:\:\:\:\:\:\:\:\:\:\:\:\:\:\:\:\:\:\:\:\:\:\:\:\:\:\:\:\:\:\:\:\:+e^{- \frac{(t_0 + |\bm L|+\ii (\Omega_1 T_1^2- \Omega_2 T_2^2))^2}{2(T_1^2 + T_2^2 + \sigma_1^2 + \sigma_2^2)}}\Bigg(\ii + \text{erfi}\left(\tfrac{t_0 + |\bm L|+\ii (\Omega_1 T_1^2- \Omega_2 T_2^2)}{\sqrt{2}\sqrt{T_1^2 + T_2^2 + \sigma_1^2 + \sigma_2^2}}\right)\Bigg)\Bigg).
\end{align}
And, as it turns out, we find that
\begin{align}
    H(f_1,f_2) &= \frac{T_1T_2e^{\ii (\Omega_1 t_1+\Omega_2 t_2)}e^{- \frac{\Omega_2^2 T_2^2}{2} - \frac{\Omega_1^2 T_1^2}{2}}}{2\sqrt{2\pi}|\bm L|\sqrt{T_1^2 + T_2^2 + \sigma_1^2 + \sigma_2^2}} \Bigg(e^{- \frac{(t_0 - |\bm L|+\ii (\Omega_1 T_1^2- \Omega_2 T_2^2))^2}{2(T_1^2 + T_2^2 + \sigma_1^2 + \sigma_2^2)}}\text{erfi}\left(\frac{|\bm L|-t_0-\ii (\Omega_1 T_1^2- \Omega_2 T_2^2)}{\sqrt{2}\sqrt{T_1^2 + T_2^2 + \sigma_1^2 + \sigma_2^2}}\right)\\
     &\:\:\:\:\:\:\:\:\:\:\:\:\:\:\:\:\:\:\:\:\:\:\:\:\:\:\:\:\:\:\:\:\:\:\:\:\:\:\:\:\:\:\:\:\:\:\:\:\:\:\:\:\:\:\:\:\:\:\:\:\:\:\:\:\:\:\:\:\:\:\:\:\:\:\:\:\:\:\:+e^{- \frac{(t_0 + |\bm L|+\ii (\Omega_1 T_1^2- \Omega_2 T_2^2))^2}{2(T_1^2 + T_2^2 + \sigma_1^2 + \sigma_2^2)}}\text{erfi}\left(\frac{|\bm L|+t_0+\ii (\Omega_1 T_1^2- \Omega_2 T_2^2)}{\sqrt{2}\sqrt{T_1^2 + T_2^2 + \sigma_1^2 + \sigma_2^2}}\right)\Bigg),\label{eq:Hbig}\nonumber
\end{align}
and
\begin{align}
    E(f_1,f_2) &= \frac{T_1T_2e^{\ii (\Omega_1 t_1+\Omega_2 t_2)}e^{- \frac{\Omega_2^2 T_2^2}{2} - \frac{\Omega_1^2 T_1^2}{2}}}{2\sqrt{2\pi}|\bm L|\sqrt{T_1^2 + T_2^2 + \sigma_1^2 + \sigma_2^2}} \Bigg(e^{- \frac{(t_0 + |\bm L|+\ii (\Omega_1 T_1^2- \Omega_2 T_2^2))^2}{2(T_1^2 + T_2^2 + \sigma_1^2 + \sigma_2^2)}}-e^{- \frac{(t_0 - |\bm L|+\ii (\Omega_1 T_1^2- \Omega_2 T_2^2))^2}{2(T_1^2 + T_2^2 + \sigma_1^2 + \sigma_2^2)}}\Bigg).
\end{align}

We now move on to the Green's functions, which will allow us to compute the Feynman propagator and the symmetric propagator. In order to compute these, we will resort to integration in spacetime of the spacetime smearing functions, using the expressions for the Green's functions of Eqs.~\eqref{eq:GR} and~\eqref{eq:GA}. Unfortunately, we will have to consider a more restricted parameter space where $\sigma_1 = \sigma_2 = \sigma$ in order to solve these integrals. For the retarded Green's function, we have
\begin{align}
    G_R(f_1,f_2) &= \int \dd V\dd V' f_1(\mf x) f_2(\mf x') G_R(\mf x,\mf x')\\
    & = \frac{1}{(2\pi)^3 \sigma^6} \int \dd t \dd t' \dd^3\bm x \dd^3 \bm x' e^{- \frac{|\bm x - \bm L_1|^2}{2\sigma^2}}e^{- \frac{|\bm x' - \bm L_2|^2}{2\sigma^2}} e^{\ii \Omega_1 t} e^{- \frac{(t-t_1)^2}{2T_1^2}}e^{\ii \Omega_2 t'} e^{- \frac{(t'-t_2)^2}{2T_2^2}}\left(- \frac{\delta(t' - t + |\bm x - \bm x'|)}{4\pi|\bm x-\bm x'|} \right)\nonumber\\
    & = -\frac{1}{2(2\pi)^4 \sigma^6} \int \dd t  \dd^3\bm x \dd^3 \bm x' e^{- \frac{|\bm x - \bm L_1|^2}{2\sigma^2}}e^{- \frac{|\bm x' - \bm L_2|^2}{2\sigma^2}} e^{\ii \Omega_1 t} e^{- \frac{(t-t_1)^2}{2T_1^2}}e^{\ii \Omega_2 (t - |\bm x - \bm x'|)} e^{- \frac{(t - |\bm x - \bm x'|-t_2)^2}{2T_2^2}}\frac{1}{|\bm x - \bm x'|}.\nonumber
\end{align}
We now perform the change of variables
\begin{align}
    \bm x = \frac{1}{\sqrt{2}}(\bm u +\bm v) + \bm L_1, \quad\quad
    \bm x' = \frac{1}{\sqrt{2}}(\bm u -\bm v) + \bm L_1,
\end{align}
so that $|\bm x - \bm x'| = \sqrt{2} |\bm v|$. Defining $\bm L = \bm L_1 - \bm L_2$, the integral becomes:
\begin{align}
    G_R(f_1,f_2) & = -\frac{1}{2(2\pi)^4 \sigma^6} \int \dd^3\bm u \dd^3 \bm v e^{- \frac{|\bm u + \bm v|^2}{4\sigma^2}}e^{- \frac{|\bm u - \bm v + \sqrt{2}\bm L|^2}{4\sigma^2}}\frac{1}{\sqrt{2}|\bm v|} \int \dd t e^{\ii \Omega_1 t} e^{- \frac{(t-t_1)^2}{2T_1^2}}e^{\ii \Omega_2 (t - \sqrt{2}|\bm v|)} e^{- \frac{(t - \sqrt{2}|\bm v|-t_2)^2}{2T_2^2}}\\
    & = -\frac{1}{2(2\pi)^4 \sigma^6} \int   \dd^3\bm u \dd^3 \bm v e^{- \frac{|\bm u|^2 + |\bm v|^2}{4\sigma^2}}e^{- \frac{|\bm u|^2 +|\bm v|^2 + 2|\bm L|^2}{4\sigma^2}}e^{-\frac{\sqrt{2}\bm u\cdot \bm L + \sqrt{2}\bm v \cdot \bm L}{2\sigma^2}} \frac{1}{\sqrt{2}|\bm v|} \nonumber\\
    &\:\:\:\:\:\:\:\:\:\:\:\:\:\:\:\:\:\:\:\:\:\:\:\:\:\:\:\:\:\:\:\:\:\:\:\:\:\:\:\:\:\:\:\:\:\:\:\:\:\:\:\:\:\:\:\:\:\:\:\:\:\:\:\:\:\:\:\:\:\:\:\:\:\:\:\:\:\:\:\:\:\:\:\:\:\:\:\:\:\:\:\:\:\:\:\:\:\:\:\:\:\:\:\:\:\times\int \dd t e^{\ii \Omega_1 t} e^{- \frac{(t-t_1)^2}{2T_1^2}}e^{\ii \Omega_2 (t - \sqrt{2}|\bm v|)} e^{- \frac{(t - \sqrt{2}|\bm v|-t_2)^2}{2T_2^2}}\nonumber\\
    &= -\frac{e^{- \frac{|\bm L|^2}{2\sigma^2}}}{2(2\pi)^4 \sigma^6} \int   \dd^3\bm u \dd^3 \bm v e^{-\frac{|\bm u|^2}{2\sigma^2}}e^{-\frac{|\bm v|^2}{2\sigma^2}} e^{-\frac{\bm u\cdot \bm L}{\sqrt{2}\sigma^2}}e^{\frac{\bm v \cdot \bm L}{\sqrt{2}\sigma^2}}\frac{1}{\sqrt{2}|\bm v|} \int \dd t e^{\ii \Omega_1 t} e^{- \frac{(t-t_1)^2}{2T_1^2}}e^{\ii \Omega_2 (t - \sqrt{2}|\bm v|)} e^{- \frac{(t - \sqrt{2}|\bm v|-t_2)^2}{2T_2^2}}.\nonumber
\end{align}
We can now proceed to first solve for the angular integrals in $\bm u$ and $\bm v$, and then the radial integral in $|\bm u|$, resulting in
\begin{align}
    G_R(f_1,f_2) & = -\frac{e^{- \frac{|\bm L|^2}{2\sigma^2}}}{2(2\pi)^2 \sigma^6} \int   \dd |\bm u| \dd |\bm v| \,|\bm u|^2 |\bm v|^2 e^{-\frac{|\bm u|^2}{2\sigma^2}}e^{-\frac{|\bm v|^2}{2\sigma^2}} \frac{2\sqrt{2} \sigma^2\sinh(\frac{|\bm L||\bm u|}{\sqrt{2}\sigma^2})}{|\bm u||\bm L|}\frac{2\sqrt{2} \sigma^2\sinh(\frac{|\bm L||\bm v|}{\sqrt{2}\sigma^2})}{|\bm v||\bm L|}\frac{1}{\sqrt{2}|\bm v|}\nonumber \\
    &\:\:\:\:\:\:\:\:\:\:\:\:\:\:\:\:\:\:\:\:\:\:\:\:\:\:\:\:\:\:\:\:\:\:\:\:\:\:\:\:\:\:\:\:\:\:\:\:\:\:\:\:\:\:\:\:\:\:\:\:\:\:\:\:\:\:\:\:\:\:\:\:\:\:\:\:\:\:\:\:\:\:\:\:\:\:\:\:\:\:\:\:\:\:\:\:\:\:\:\:\:\:\:\:\:\times\int \dd t e^{\ii \Omega_1 t} e^{- \frac{(t-t_1)^2}{2T_1^2}}e^{\ii \Omega_2 (t - \sqrt{2}|\bm v|)} e^{- \frac{(t - \sqrt{2}|\bm v|-t_2)^2}{2T_2^2}}\nonumber\\
    & = -\frac{e^{- \frac{|\bm L|^2}{2\sigma^2}}}{\sqrt{2}\pi^2\sigma^2|\bm L|^2} \int  \dd |\bm u| \dd |\bm v| \,|\bm u| e^{-\frac{|\bm u|^2}{2\sigma^2}} \sinh(\frac{|\bm L||\bm u|}{\sqrt{2}\sigma^2})e^{-\frac{|\bm v|^2}{2\sigma^2}}\sinh(\frac{|\bm L||\bm v|}{\sqrt{2}\sigma^2})\nonumber\\
    &\:\:\:\:\:\:\:\:\:\:\:\:\:\:\:\:\:\:\:\:\:\:\:\:\:\:\:\:\:\:\:\:\:\:\:\:\:\:\:\:\:\:\:\:\:\:\:\:\:\:\:\:\:\:\:\:\:\:\:\:\:\:\:\:\:\:\:\:\:\:\:\:\:\:\:\:\:\:\:\:\:\:\:\:\:\:\:\:\:\:\:\:\:\:\:\:\:\:\:\:\:\:\:\:\:\times \int \dd t e^{\ii \Omega_1 t} e^{- \frac{(t-t_1)^2}{2T_1^2}}e^{\ii \Omega_2 (t - \sqrt{2}|\bm v|)} e^{- \frac{(t - \sqrt{2}|\bm v|-t_2)^2}{2T_2^2}}\nonumber\\
    & = -\frac{e^{- \frac{|\bm L|^2}{2\sigma^2}}}{\sqrt{2}\pi^2\sigma^2|\bm L|^2} \int  \dd |\bm v| \, \frac{\sqrt{\pi}\sigma |\bm L|}{2}e^{\frac{|\bm L|^2}{4 \sigma^2}}e^{-\frac{|\bm v|^2}{2\sigma^2}} \sinh(\frac{|\bm L||\bm v|}{\sqrt{2}\sigma^2}) \int \dd te^{\ii \Omega_1 t} e^{- \frac{(t-t_1)^2}{2T_1^2}}e^{\ii \Omega_2 (t - \sqrt{2}|\bm v|)} e^{- \frac{(t - \sqrt{2}|\bm v|-t_2)^2}{2T_2^2}}\nonumber\\
    & = -\frac{e^{- \frac{|\bm L|^2}{4\sigma^2}}}{2\sqrt{2}\pi^{3/2}\sigma|\bm L|} \int \dd |\bm v| \, e^{-\frac{|\bm v|^2}{2\sigma^2}} \sinh(\frac{|\bm L||\bm v|}{\sqrt{2}\sigma^2}) \int \dd t e^{\ii \Omega_1 t} e^{- \frac{(t-t_1)^2}{2T_1^2}}e^{\ii \Omega_2 (t - \sqrt{2}|\bm v|)} e^{- \frac{(t - \sqrt{2}|\bm v|-t_2)^2}{2T_2^2}}.\label{eq:GRfinalInt}
\end{align}
Defining $t_0 = t_1 - t_2$, the integral in $t$ yields
\begin{align}
     \int \dd t e^{\ii \Omega_1 t} e^{- \frac{(t-t_1)^2}{2T_1^2}}e^{\ii \Omega_2 (t - \sqrt{2}|\bm v|)}& e^{- \frac{(t - \sqrt{2}|\bm v|-t_2)^2}{2T_2^2}}\nonumber\\
     &\!\!\!\! = \frac{\sqrt{2\pi}T_1 T_2}{\sqrt{T_1^2 + T_2^2}}e^{- \frac{|\bm v|^2}{T_1^2 + T_2^2}}e^{\frac{\sqrt{2}|\bm v|(t_0- \ii (\Omega_1 T_1^2 - \Omega_2 T_2^2))}{(T_1^2 + T_2^2)}}e^{-\frac{t_0^2}{2(T_1^2 + T_2^2)}-\frac{T_1^2T_2^2(\Omega_1+\Omega_2)^2}{2(T_1^2 + T_2^2)}+\ii(\Omega_1 + \Omega_2)\frac{(t_1 T_2^2 + t_2 T_1^2)}{T_1^2 + T_2^2}}.
\end{align}
The final integral in $|\bm v|$ is merely a combination of Gaussian integrals. After simplifications, we finally find
\begin{align}
    G_R(f_1,f_2) = - \frac{T_1T_2e^{\ii (\Omega_1 t_1+\Omega_2 t_2)}e^{- \frac{\Omega_2^2 T_2^2}{2} - \frac{\Omega_1^2 T_1^2}{2}}}{4\sqrt{2\pi}|\bm L|\sqrt{T_1^2 + T_2^2 + 2\sigma^2}}\Bigg(&e^{- \frac{(t_0 - |\bm L|+\ii (\Omega_1 T_1^2- \Omega_2 T_2^2))^2}{2(T_1^2 + T_2^2 + 2\sigma^2)}}\left(1 + \text{erf}\left(\tfrac{|\bm L|(T_1^2+T_2^2)+2\sigma^2(t_0+\ii (\Omega_1 T_1^2- \Omega_2 T_2^2))}{2\sigma\sqrt{T_1^2 + T_2^2}\sqrt{T_1^2 + T_2^2 + 2\sigma^2}}\right)\right)\nonumber\\
    &\!\!\!\!\!\!\!\!\!\!\!\!\!\!\!\!\!\!+e^{- \frac{(t_0 + |\bm L|+\ii (\Omega_1 T_1^2- \Omega_2 T_2^2))^2}{2(T_1^2 + T_2^2 + 2\sigma^2)}}\left(-1+\text{erf}\left(\tfrac{|\bm L|(T_1^2+T_2^2)-2\sigma^2(t_0+\ii (\Omega_1 T_1^2- \Omega_2 T_2^2))}{2\sigma\sqrt{T_1^2 + T_2^2}\sqrt{T_1^2 + T_2^2 + 2\sigma^2}}\right)\right)\Bigg).\label{eq:GRbig}
\end{align}
Combining $G_R(f_1,f_2)$ and $E(f_1,f_2)$, we can then find $G_A(f_1,f_2)$,
\begin{align}
    G_A(f_1,f_2) = - \frac{T_1T_2e^{\ii (\Omega_1 t_1+\Omega_2 t_2)}e^{- \frac{\Omega_2^2 T_2^2}{2} - \frac{\Omega_1^2 T_1^2}{2}}}{4\sqrt{2\pi}|\bm L|\sqrt{T_1^2 + T_2^2 + 2\sigma^2}}\Bigg(&e^{- \frac{(t_0 - |\bm L|+\ii (\Omega_1 T_1^2- \Omega_2 T_2^2))^2}{2(T_1^2 + T_2^2 + 2\sigma^2)}}\left(-1 + \text{erf}\left(\tfrac{|\bm L|(T_1^2+T_2^2)+2\sigma^2(t_0+\ii (\Omega_1 T_1^2- \Omega_2 T_2^2))}{2\sigma\sqrt{T_1^2 + T_2^2}\sqrt{T_1^2 + T_2^2 + 2\sigma^2}}\right)\right)\nonumber\\
    &\!\!\!\!\!\!\!\!\!\!\!\!\!\!\!\!\!\!+e^{- \frac{(t_0 + |\bm L|+\ii (\Omega_1 T_1^2- \Omega_2 T_2^2))^2}{2(T_1^2 + T_2^2 + 2\sigma^2)}}\left(1+\text{erf}\left(\tfrac{|\bm L|(T_1^2+T_2^2)-2\sigma^2(t_0+\ii (\Omega_1 T_1^2- \Omega_2 T_2^2))}{2\sigma\sqrt{T_1^2 + T_2^2}\sqrt{T_1^2 + T_2^2 + 2\sigma^2}}\right)\right)\Bigg).\label{eq:GAbig}
\end{align}
Finally, we find $\Delta(f_1,f_2)$ by adding Eqs.~\eqref{eq:GRbig} and~\eqref{eq:GAbig}:
\begin{align}
    \Delta(f_1,f_2) = - \frac{T_1T_2e^{\ii (\Omega_1 t_1+\Omega_2 t_2)}e^{- \frac{\Omega_2^2 T_2^2}{2} - \frac{\Omega_1^2 T_1^2}{2}}}{2\sqrt{2\pi}|\bm L|\sqrt{T_1^2 + T_2^2 + 2\sigma^2}}\Bigg(&e^{- \frac{(t_0 - |\bm L|+\ii (\Omega_1 T_1^2- \Omega_2 T_2^2))^2}{2(T_1^2 + T_2^2 + 2\sigma^2)}} \text{erf}\left(\tfrac{|\bm L|(T_1^2+T_2^2)+2\sigma^2(t_0+\ii (\Omega_1 T_1^2- \Omega_2 T_2^2))}{2\sigma\sqrt{T_1^2 + T_2^2}\sqrt{T_1^2 + T_2^2 + 2\sigma^2}}\right)\label{eq:Deltabig}\\
    &\:\:\:\:\:\:\:\:\:\:\:+e^{- \frac{(t_0 + |\bm L|+\ii (\Omega_1 T_1^2- \Omega_2 T_2^2))^2}{2(T_1^2 + T_2^2 + 2\sigma^2)}}\text{erf}\left(\tfrac{|\bm L|(T_1^2+T_2^2) - 2\sigma^2(t_0+\ii (\Omega_1 T_1^2- \Omega_2 T_2^2))}{2\sigma\sqrt{T_1^2 + T_2^2}\sqrt{T_1^2 + T_2^2 + 2\sigma^2}}\right)\Bigg).\nonumber
\end{align}
Using Eqs.~\eqref{eq:Hbig} and~\eqref{eq:Deltabig}, one can then find $G_F(f_1,f_2) = \frac{1}{2}H(f_1,f_2) + \frac{\ii}{2}\Delta(f_1,f_2)$.

In order to obtain the field bi-ditributions evaluated at the field's momentum $\hat{\pi}(\mf x) = \partial_t \hat{\phi}(\mf x)$ rather than at the field's amplitude smeared in Gaussian spacetime regions, one can simply differentiate the results of this section with respect to the parameters $\Omega_1$, $\Omega_2$, using
\begin{equation}
    \omega( \hat{\pi}(f_1)\hat{\phi}(f_2)) = \omega( \hat{\phi}(-\partial_tf_1)\hat{\phi}(f_2)),
\end{equation}
and noticing that for the functions used here
\begin{equation}
    \partial_t f_1(\mf x) = \left(- \frac{t}{T_1^2} + \frac{t_1}{T_1^2} + \ii \Omega_1\right) f_1(\mf x) = \frac{\ii}{T_1^2}\dv{}{\Omega_1}f_1(\mf x) + \left(\frac{t_1}{T_1^2} + \ii \Omega_1\right) f_1(\mf x).
\end{equation}
That is, one finds that
\begin{equation}
    \omega( \hat{\pi}(f_1)\hat{\phi}(f_2)) =  - \frac{\ii}{T_1^2}\dv{}{\Omega_1}W(f_1,f_2) -\left(\frac{t_1}{T_1^2} + \ii \Omega_1\right) W(f_1,f_2).
\end{equation}
Analogous expressions are valid for higher derivatives of the field in both arguments, and for the other bi-distributions $H$, $E$, $G_R$, $G_A$, $G_F$ and $\Delta$. For brevity, we will not write these explicit expressions here, but they can be straightforwardly computed.

\twocolumngrid

\bibliography{references}

\end{document}